\documentclass[english,reprint,amsmath,amssymb,aps,pra,superscriptaddress]{revtex4-1}
\usepackage[latin9]{inputenc}
\usepackage{babel}
\usepackage{amsmath}
\usepackage{graphicx}
\usepackage{esint}
\usepackage[unicode=true]
 {hyperref}
\usepackage{breakurl}

\makeatletter

\newcommand*\LyXThinSpace{\,\hspace{0pt}}

\usepackage{dcolumn}% Align table columns on decimal point
\usepackage{bm}% bold math

\makeatother

\begin{document}

\title{Three-body bound states of an atom in a Fermi mixture}

\author{Ali Sanayei}
\email{asanayei@physnet.uni-hamburg.de}

\selectlanguage{english}%

\address{Zentrum f{\"u}r Optische Quantentechnologien, Universit{\"a}t Hamburg,
Luruper Chaussee 149, D-22761 Hamburg, Germany}

\address{Institut f{\"u}r Laserphysik, Universit{\"a}t Hamburg, Luruper
Chaussee 149, D-22761 Hamburg, Germany}

\author{Ludwig Mathey}
\email{lmathey@physnet.uni-hamburg.de}

\selectlanguage{english}%

\address{Zentrum f{\"u}r Optische Quantentechnologien, Universit{\"a}t Hamburg,
Luruper Chaussee 149, D-22761 Hamburg, Germany}

\address{Institut f{\"u}r Laserphysik, Universit{\"a}t Hamburg, Luruper
Chaussee 149, D-22761 Hamburg, Germany}

\address{The Hamburg Centre for Ultrafast Imaging, Luruper Chaussee 149, D-22761
Hamburg, Germany}

\date{\today}
\begin{abstract}
We determine the three-body bound states of an atom in a Fermi mixture.
Compared to the Efimov spectrum of three atoms in vacuum, we show
that the Fermi seas deform the Efimov spectrum systematically. We
demonstrate that this effect is more pronounced near unitarity, for
which we give an analytical estimate. We show that in the presence
of Fermi seas, the three-body bound states obey a generalized discrete
scaling law. For an experimental confirmation of our prediction, we
propose three signatures of three-body bound states of an ultracold
Fermi mixture of Yb isotopes, and provide an estimate for the onset
of the bound state and the binding energy.
\end{abstract}
\maketitle

\section{introduction \label{Sec_I}}

In a seminal paper, Ref. \cite{Efimov_1}, Efimov showed that three
bosons that interact attractively in vacuum via short-range interactions
form three-body bound states at interaction strengths that are not
yet sufficient to support two-body bound states. He also showed that
the number of the three-body bound states is in principle infinite,
and that there is a geometric scaling law that governs the bound states
\cite{Efimov_2,Braaten_universality,Grimm_and_Ferlaino,Pascal_review_paper,Greene_review,DIncao_review}.
Technical advances in the trapping and cooling of atoms \cite{trapping_atoms,kF_and_n}
as well as in the Feshbach resonances \cite{many_body_ultracold_atoms,Feshbach_resonance_Review}
have led to the observation of the Efimov effect in ultracold atomic
gases \cite{Efimov_1st_expt,expt_Li_K,expt_ultracold_second_trimer,expt_Li_Cs_1,expt_Li_Cs_2,Efimov_ultracold_summary}
and helium beams experiments \cite{expt_helium_beam_1,expt_helium_beam_2}.
Excited three-body bound states were observed \cite{expt_ultracold_second_trimer,expt_Efimov_scaling_factor},
and the Efimov scaling law was confirmed. The Efimov effect was also
generalized to more than three particles \cite{Pascal_review_paper,four_body_0}.
It was shown that for a critical mass ratio three fermions and a lighter
particle form a four-body bound state \cite{four_body_1}. The four-body
bound states of two heavy and two light bosons for different mass
ratios was investigated in Ref. \cite{four_body_2}. The formation
of a five-body bound state in fermionic mixtures was discussed in
Ref. \cite{five_body_1}.

Recently, we demonstrated the formation of three-electron bound states
in conventional superconductors, and showed that the trimer state
competes with the formation of the two-electron Cooper pair \cite{Sanayei}.
For that, we modeled the interaction between two particles ``$i$''
and ``$j$'' as a negative constant $g_{ij}$ in momentum space
for an incoming and outgoing momentum of a particle smaller than a
cutoff $\Lambda_{i}$, following the reasoning of the Cooper problem
\cite{Cooper}. We fixed the cutoffs by a typical value of the Debye
energy in a conventional superconductor \cite{Sanayei}. In this paper
we determine the three-body bound states of an atom in a Fermi mixture
for contact interactions. To describe contact interactions we take
the limit of the cutoffs $\Lambda_{i}$ to infinity. We show that
this model is separable \cite{separable}, leading to a system of
two coupled integral equations. This model enables us to calculate
the three-body bound-state spectrum in the presence of Fermi seas. 

\begin{figure}[t]
\includegraphics[width=0.96\columnwidth]{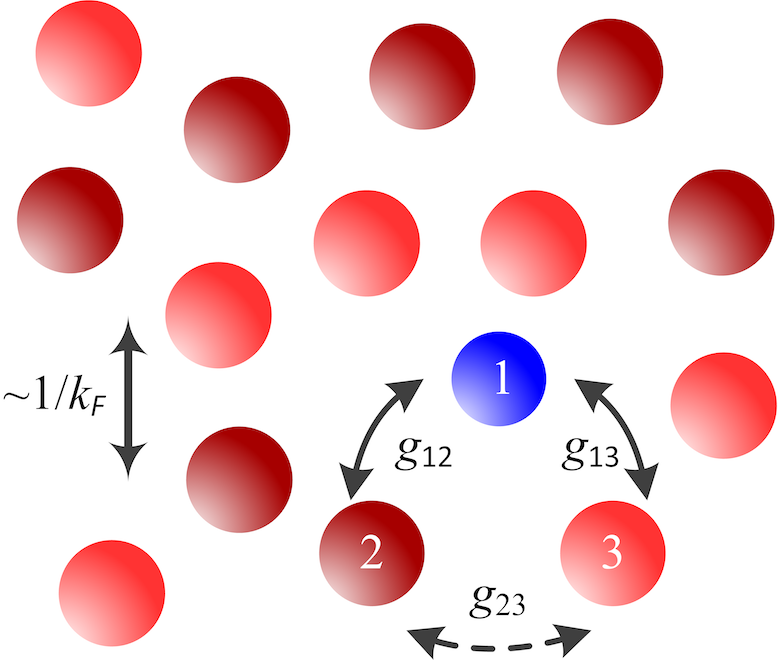}

\caption{Sketch of an atom in a Fermi mixture. All species interact attractively
via contact interactions. Species ``2'' and ``3'' are a Fermi
mixture, and atom ``1'' can in general be a boson or fermion. The
interaction strengths are shown by three negative constants $g_{12}$,
$g_{13}$, and $g_{23}$. The species ``2'' and ``3'' are assumed
to be in different internal states and $m_{3}=m_{2}$. The density
of each species ``2'' and ``3'' is $n_{\mathrm{tot}}/2$, defining
an inert Fermi sea with the Fermi momentum $k_{F}=(3\pi^{2}n_{\mathrm{tot}})^{1/3}$.
The interatomic distances are proportional to $1/k_{F}$. }

\label{Fig1}
\end{figure}

In this work, we consider a cold-atom system of Fermi mixtures. We
assume a density of the species, labeled ``2'', that interacts attractively
with another species of the same density, labeled ``3''. We assume
that the two species ``2'' and ``3'' are in different internal
states. Next, we include an additional atom, labeled ``1'', that
interacts attractively with the other atoms via contact interactions;
see Fig. \ref{Fig1}. In general, the three masses $m_{1}$, $m_{2}$,
and $m_{3}$ can be different, but we are primarily interested in
the case $m_{3}=m_{2}$. We assume that atom ``1'' is a fermion.
A similar analysis can be applied when it is a boson. The species
``2'' and ``3''define the Fermi seas with the Fermi momentum $k_{F}$.
This imposes the constraints $k_{2}>k_{F}$ and $k_{3}>k_{F}$ on
the momentum of atoms ``2'' and ``3'', respectively. We also assume
that the interatomic distances, proportional to $1/k_{F}$, are much
larger than the range of the atomic interactions. With this, we neglect
the many-body effects on the formation of a three-body bound state
within the interatomic distances. For contact interactions we introduce
the \emph{s}-wave scattering lengths as it relates to the contact
interaction in its regularized form. We also define a three-body parameter,
$\Lambda$, in order to regularize the range of the three-body interactions
and to prevent Thomas collapse \cite{Thomas_collapse}. This parameter
defines a length scale of the range of the atomic interactions using
the van der Waals length \cite{Pascal_review_paper,three_body_parameter_1,three_body_parameter_2}. 

We calculate the three-body bound states for different mass ratios.
We provide an analytical description of the lowest-energy two-body
bound states and the two-body continuum, and find the three-body bound-state
solutions numerically. For a noninteracting mixture, $g_{23}=0$,
we provide an analytical formula for the onset of the lowest-energy
two-body bound state at zero energy. For a high mass ratio $m_{2}/m_{1}$,
where the excited three-body bound states appear, we also find an
analytical estimate for the onset of a highest-energy excited three-body
bound state at zero energy. With this, we can estimate the amount
of the shift that the spectrum undergoes near unitarity due to the
Fermi seas. Further, for our system and interaction model we demonstrate
that a generalized scaling law governs the three-body bound states
in the presence of Fermi seas. Finally, we propose three experimental
scenarios in an ultracold system of fermionic mixtures of Yb isotopes
to observe three-body bound states in the presence of Fermi seas.
Here the $^{171}\mathrm{Yb}$ isotopes, that are in two different
internal states, constitute the Fermi seas, and interact attractively
with $^{173}\mathrm{Yb}$. We predict the onset of the three-body
bound states and provide an estimate for the threshold energy.

This paper is organized as follows. In Sec. \ref{Sec_II} we provide
the main formulation of the problem for contact interactions, and
derive a system of two coupled integral equations describing an atom
in a Fermi mixture. In Sec. \ref{Sec_III} we represent our results
for two- and three interacting pairs in the presence of Fermi seas,
and demonstrate a generalized scaling law governing the three-body
bound states. Here we also derive an analytical estimate to describe
the effect of the Fermi seas near unitarity. In Sec. \ref{Sec_IV}
we present three experimental signatures of a three-body bound state
in an ultracold Fermi mixture of Yb isotopes. Finally, in Sec. \ref{Sec_V}
we present our concluding remarks.

\section{formulation of the problem\label{Sec_II}}

The Schr{\"o}dinger equation for a system of three atoms in momentum
space is

\begin{equation}
\left(\frac{\hbar^{2}k_{1}^{2}}{2m_{1}}+\frac{\hbar^{2}k_{2}^{2}}{2m_{2}}+\frac{\hbar^{2}k_{3}^{2}}{2m_{3}}+\hat{U}_{12}+\hat{U}_{13}+\hat{U}_{23}-E\right)\psi=0,\label{main Schrodinger Eq}
\end{equation}
where $\hbar$ is the reduced Planck's constant, $m_{i}$ and $\mathbf{k}_{i}$
is the atom mass and momentum, respectively, $E$ is the energy, and
$\psi=\psi(\mathbf{k}_{1},\mathbf{k}_{2},\mathbf{k}_{3})$ is the
wave function. We consider the interaction $\hat{U}_{ij}$ between
the atom ``$i$'' and ``$j$'', $i,j=1,2,3$ and $i\neq j$, as

\begin{equation}
\hat{U}_{ij}\psi=g_{ij}\theta_{\Lambda_{i}}(\mathbf{k}_{i})\theta_{\Lambda_{j}}(\mathbf{k}_{j})\int\frac{d^{3}\mathbf{q}}{(2\pi)^{3}}\theta_{\Lambda_{i}}(\mathbf{k}_{i}-\mathbf{q})\theta_{\Lambda_{j}}(\mathbf{k}_{j}+\mathbf{q})\psi,\label{Cooper interaction operator}
\end{equation}
where $\mathbf{q}$ is the momentum transfer \cite{momentum_transfer}
and $g_{ij}<0$ is the interaction strength; see Ref. \cite{Sanayei}.
The resulting operators $\hat{U}_{ij}\psi$ are given in Appendix
B. The cutoff function $\theta_{a,b}(\mathbf{k})$ for two real numbers
$0\leqslant a<b$ is defined as

\begin{equation}
\theta_{a,b}(\mathbf{k})=\begin{cases}
1 & \text{for }a\leqslant|\mathbf{k}|\leqslant b,\\
0 & \text{otherwise},
\end{cases}\label{cutoff func}
\end{equation}
and $\theta_{b}(\mathbf{k})\equiv\theta_{0,b}(\mathbf{k})$. Here
we consider three-body bound states with vanishing total momentum.
We also consider a singlet state for the species \textquotedblleft 2\textquotedblright{}
and \textquotedblleft 3\textquotedblright{} in the following. The
Fermi seas demand the constraints $k_{2}>k_{F}$ and $k_{3}>k_{F}$
on the momentum of the atoms ``2'' and ``3'', respectively. The
threshold energy of the bound states is 
\begin{figure}[t]
\includegraphics[width=1\columnwidth]{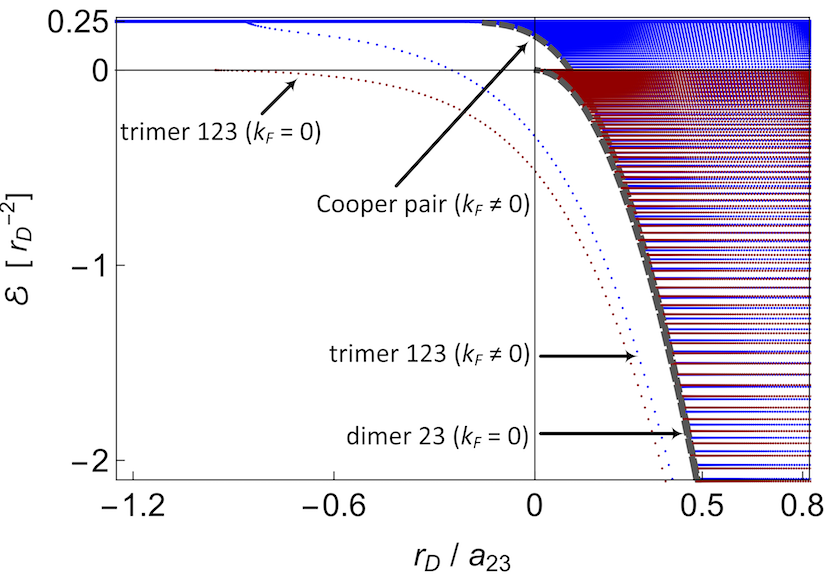}

\caption{Energy $\mathcal{E}=2\mu_{12}E/\hbar^{2}$ in units of $r_{D}^{-2}$
vs $r_{D}/a_{23}$ for three interacting pairs, where $m_{2}/m_{1}=1$
and $a_{12}\approx-36r_{D}$. The red curves show the solution in
vacuum, $k_{F}=0$. The blue curves show the result in the presence
of Fermi seas, $k_{F}r_{D}\approx0.17$. The single blue curve is
the three-body bound-state solution for $k_{F}\protect\neq0$. The
gray dashed curves are the lowest-energy two-body bound-state solutions
of the two-body continuum in vacuum, cf. Eq. (\ref{first dimer in vacuum}),
and in the presence of Fermi seas; cf. Eq. (\ref{Cooper pair solution}).
The onset of the two-body bound-state continuum is shifted towards
negative values of $a_{23}$. The onset of the three-body bound state
is pushed towards positive values of $a_{23}$. The dependence of
the trimer energy on $a_{23}$ is modified noticeably.}

\label{Fig2}
\end{figure}

\begin{equation}
E_{\mathrm{thr}}=\frac{\hbar^{2}}{m_{2}}k_{F}^{2}=2E_{F},\label{threshold energy}
\end{equation}
where $E_{F}$ denotes the Fermi energy and $m_{3}=m_{2}$. To describe
contact interactions we take the limit of the cutoffs $\Lambda_{i}$
and $\Lambda_{j}$ to infinity. We introduce the \emph{s}-wave scattering
length, $a_{ij}$, using the following regularization identity:

\begin{equation}
\frac{2\pi\hbar^{2}}{\mu_{ij}}\frac{1}{g_{ij}}+\frac{2}{\pi}\Lambda_{j}\equiv\frac{1}{a_{ij}}\text{ as }\Lambda_{j}\rightarrow\infty,\label{s wave scattering length}
\end{equation}
for $i,j=1,2,3$ and $i\neq j$; see Appendix A. Here, $\mu_{ij}$
is a reduced mass, $1/\mu_{ij}=1/m_{i}+1/m_{j}$, $m_{3}=m_{2}$,
and $\Lambda_{i}\sim\Lambda_{j}$. Next, we define $\Lambda$ as the
three-body parameter that fixes the range of the atomic interactions
and regularizes the three-body bound states \cite{Pascal_review_paper,three_body_parameter_1,three_body_parameter_2}.
We also define a length scale, $r_{D}$, as 

\begin{equation}
r_{D}=\frac{1}{\Lambda}.\label{length scale}
\end{equation}
The value of $\Lambda$ is chosen such that $\Lambda\gg k_{F}$, implying
that $r_{D}\ll1/k_{F}$. With this, we neglect the many-body effects
on the formation of a three-body bound state. We determine $r_{D}$
as the range of the atomic interactions, which we take as the van
der Waals length, $\ell_{ij}^{(\mathrm{vdW})}=\frac{1}{2}(2\mu_{ij}C_{6}/\hbar^{2})^{1/4}$,
where $C_{6}$ is a dispersive coefficient associated with the polarizability
of the electronic cloud of the atoms \cite{Pascal_review_paper,Feshbach_resonance_Review,C6_coeff_1,C6_coeff_Yb_1,C6_coeff_2,C6_coeff_3}.
We also assume that the range of the interactions is much larger than
the Compton wave length of the particles, $r_{D}\gg\lambda_{\mathrm{C}}$,
implying that relativistic corrections to the three-body bound-state
spectrum can be neglected. In what follows, we refer to a two-body
bound state of atoms ``$i$'' and ``$j$'' as a dimer-$ij$, and
to a three-body bound state of atoms ``$i$'', ``$j$'' and ``$l$''
as a trimer-$ijl$. We also refer to a two-body bound state of species
``2'' and ``3'' as a Cooper pair for $k_{F}\neq0$, and as a dimer-23
for $k_{F}=0$.

\begin{figure}[t]
\includegraphics[width=1\columnwidth]{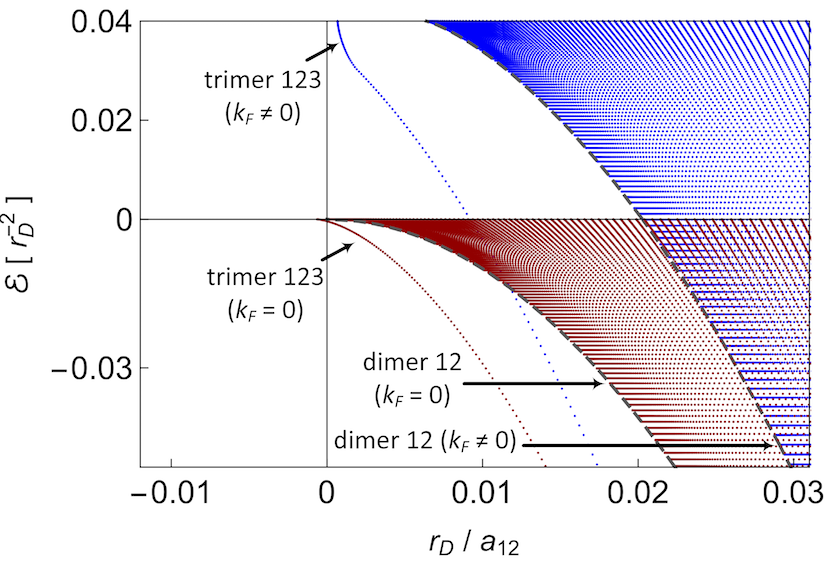}

\caption{Energy $\mathcal{E}=2\mu_{12}E/\hbar^{2}$ in units of $r_{D}^{-2}$
vs $r_{D}/a_{12}$ for $g_{23}=0$ and $m_{2}/m_{1}=1$. The single
red curve is the three-body bound-state solution for $k_{F}=0$, and
the single blue curve is the solution for $k_{F}r_{D}\approx0.02$.
The gray dashed curves are the lowest-energy two-body bound states
of the two-body continuum in vacuum, cf. Eq. (\ref{first dimer in vacuum}),
and in the presence of Fermi seas; cf. Eq. (\ref{solution of dimers-12}).
The Fermi seas push the onset of the two-body bound-state continuum
as well as the onset of the three-body bound state to positive values
of $a_{12}$.}

\label{Fig3}
\end{figure}

\begin{figure*}
\includegraphics[width=0.5\textwidth]{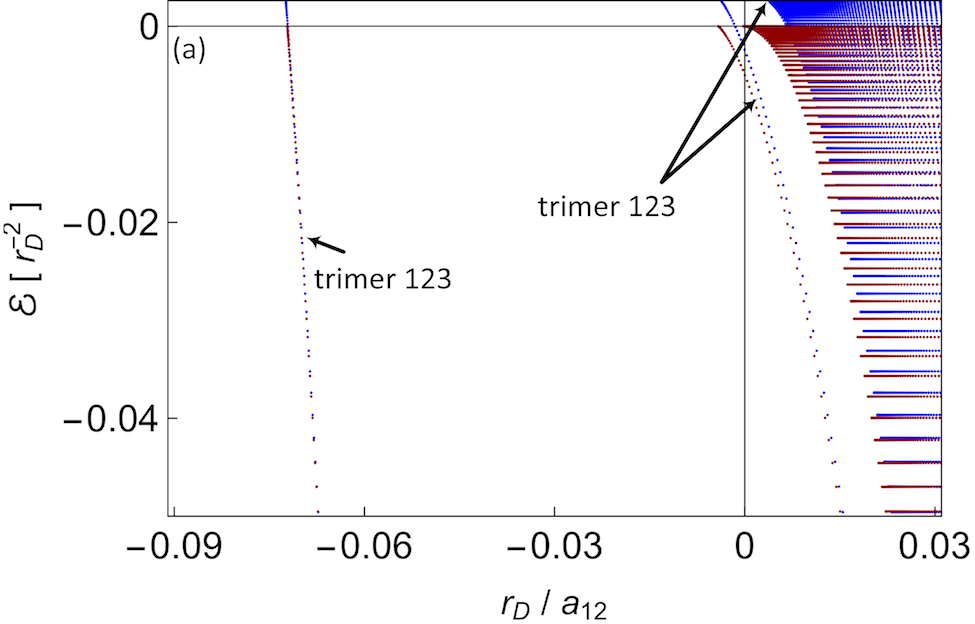}$\;$\includegraphics[width=0.47\textwidth]{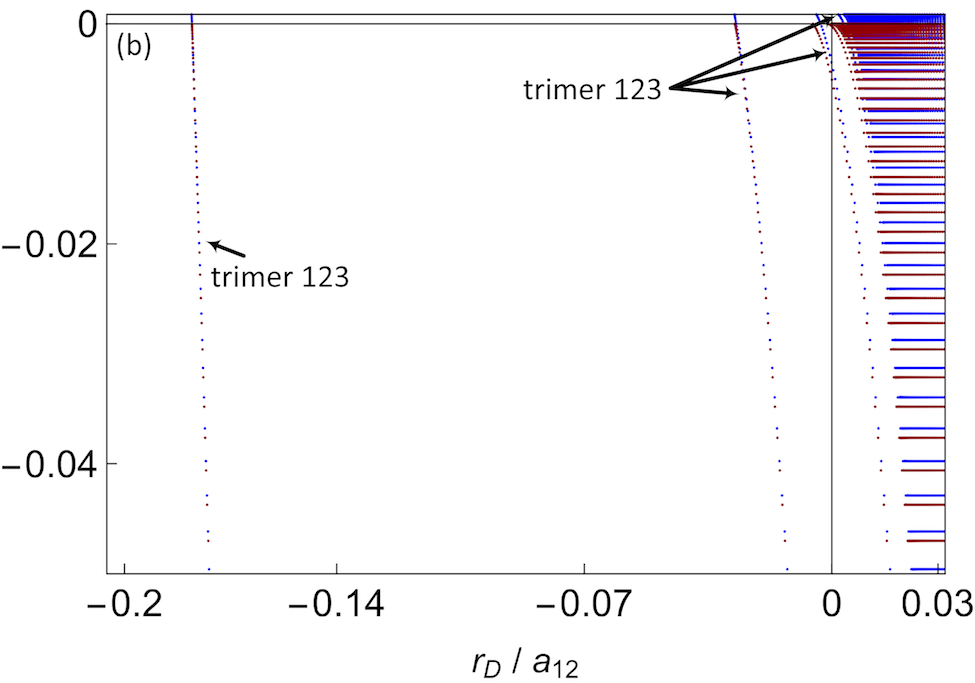}

\caption{Energy $\mathcal{E}=2\mu_{12}E/\hbar^{2}$ in units of $r_{D}^{-2}$
vs $r_{D}/a_{12}$ for $g_{23}=0$. Red curves correspond to Efimov
states, $k_{F}=0$, and blue curves are the results for $k_{F}r_{D}\approx0.01$:
(a) $m_{2}/m_{1}\approx6.64$, (b) $m_{2}/m_{1}\approx22.26$. As
the mass ratio $m_{2}/m_{1}$ increases, excited three-body bound
states appear. A zoom on the region where a highest-energy excited
three-body bound state emerges is depicted in Fig. \ref{Fig5}.}

\label{Fig4}
\end{figure*}

We note that the interaction model (\ref{Cooper interaction operator})
is separable, as shown in Appendix B. This constitutes a system of
the two coupled integral equations of the functions $F_{1}$ and $F_{2}$:

\begin{align}
\Omega_{12}(g_{12},\mathbf{k}_{2};k_{F},E)F_{2}(\mathbf{k}_{2})= & \xi_{1}(\mathbf{k}_{2};F_{2})+\xi_{2}(\mathbf{k}_{2};F_{1}),\label{Main Eq 1}\\
\Omega_{23}(g_{23},\mathbf{k}_{1};k_{F},E)F_{1}(\mathbf{k}_{1})= & \xi_{3}(\mathbf{k}_{1};F_{2}).\label{Main Eq 2}
\end{align}
The two functions $\Omega_{12}$ and $\Omega_{23}$ describe the two-body
bound state continuum, dimers-12 and dimers-23, respectively:

\begin{equation}
\Omega_{12}(g_{12},\mathbf{k}_{2};k_{F},E)=\frac{1}{g_{12}}+\int\frac{d^{3}\mathbf{p}_{3}}{(2\pi)^{3}}\,K_{1}(\mathbf{k}_{2},\mathbf{p}_{3};E),\label{Omega_12 main definition}
\end{equation}

\begin{equation}
\Omega_{23}(g_{23},\mathbf{k}_{1};k_{F},E)=\frac{1}{g_{23}}+\int\frac{d^{3}\mathbf{p}_{3}}{(2\pi)^{3}}\,K_{3}(\mathbf{k}_{1},\mathbf{p}_{3};E),\label{Omega_23 main definition}
\end{equation}
where for contact interactions we use the regularization relation
(\ref{s wave scattering length}) to introduce the \emph{s}-wave scattering
lengths. The three functions $\xi_{1}$, $\xi_{2}$, and $\xi_{3}$
describe the coupling of a pair to the third atom within the range
of the length scale $r_{D}$ that is introduced by the three-body
parameter $\Lambda$:

\begin{equation}
\xi_{1}(\mathbf{k}_{2};F_{2})=-\int\frac{d^{3}\tilde{\mathbf{p}}_{3}}{(2\pi)^{3}}\,\tilde{K}_{1}(\mathbf{k}_{2},\tilde{\mathbf{p}}_{3};E)F_{2}(\tilde{\mathbf{p}}_{3}),\label{ksi_1}
\end{equation}

\begin{equation}
\xi_{2}(\mathbf{k}_{2};F_{1})=-\int\frac{d^{3}\tilde{\mathbf{p}}_{1}}{(2\pi)^{3}}\,\tilde{K}_{2}(\mathbf{k}_{2},\tilde{\mathbf{p}}_{1};E)F_{1}(\tilde{\mathbf{p}}_{1}),\label{ksi_2}
\end{equation}

\begin{equation}
\xi_{3}(\mathbf{k}_{1};F_{2})=-2\int\frac{d^{3}\tilde{\mathbf{p}}_{3}}{(2\pi)^{3}}\,\tilde{K}_{3}(\mathbf{k}_{1},\tilde{\mathbf{p}}_{3};E)F_{2}(\tilde{\mathbf{p}}_{3});\label{ksi_3}
\end{equation}
see Appendix B. The integral kernels $K_{i}$ and $\tilde{K}_{i}$,
$i=1,2,3$, and also the functions $F_{1}$ and $F_{2}$ are represented
in Appendix B. We assume that $F_{i}(\mathbf{k})=F_{i}(k)$, implying
\emph{s}-wave symmetry of the states. We notice that the system of
the integral Eqs. (\ref{Main Eq 1}) and (\ref{Main Eq 2}) can be
interpreted as the Skorniakov\textendash Ter-Martirosian equation
for the zero-range limit of the interaction model (\ref{Cooper interaction operator});
cf. Ref. \cite{Skorniakov_Ter-Martirosian}.

\section{results\label{Sec_III}}

The coupled integral equations (\ref{Main Eq 1}) and (\ref{Main Eq 2})
describe three interacting pairs. For contact interactions and\emph{
s}-wave symmetry of the states we calculate the two functions $\Omega_{23}$
and $\Omega_{12}$ analytically; see Appendices C and D. These functions
describe the lowest-energy two-body bound states and the two-body
continuum, dimers-23 and dimers-12, respectively. Next, for a given
value of the three-body parameter $\Lambda\gg k_{F}$ we evaluate
the functions $\xi_{1}$, $\xi_{2}$, and $\xi_{3}$ numerically,
and solve the system of the integral Eqs. (\ref{Main Eq 1}) and (\ref{Main Eq 2})
in order to find the three-body bound-state solutions. For that, we
discretize the interval $(k_{F},\Lambda)$, and evaluate each integral
as a truncated sum following the Gauss-Legendre quadrature rule \cite{Gaussian_quadrature_1,Gaussian_quadrature_2,Gaussian_quadrature_3,Gaussian_quadrature_4}.
We construct the corresponding matrix equation and calculate the eigenvalues
for different values of energy $E\leqslant E_{\mathrm{thr}}$, resulting
in the \emph{s}-wave scattering lengths $a_{23}$ and $a_{12}$. We
find the values of the functions $F_{1}$ and $F_{2}$ at the grid
points as the corresponding eigenvectors; see Appendix E. We note
that the two-body bound states appear as continuum states, whereas
the three-body bound states appear at discrete energy levels.

For three interacting pairs and for a fixed value of $a_{12}$, Fig.
\ref{Fig2} shows the energy as a function of the inverse \emph{s}-wave
scattering length $1/a_{23}$ for $m_{2}/m_{1}=1$, and comparison
with the result for $k_{F}=0$. It reveals a deformation of the Efimov
spectrum in the presence of Fermi seas. We notice that for vanishing
$k_{F}$, the two-body bound-state continuum emerges at unitarity,
$a_{23}\rightarrow\pm\infty$, whereas the presence of Fermi seas
expands the region of the two-body bound states to negative values
of $a_{23}$. The single red and blue curves show the three-body bound-state
solution for $k_{F}=0$ and $k_{F}\neq0$, respectively. For $k_{F}\neq0$
the three-body bound state emerges at a larger value of $|a_{23}|$
at $E=E_{\mathrm{thr}}$, and converges asymptotically to the three-body
bound-state solution in vacuum. As a general tendency, the effect
of the Fermi seas is more pronounced as we approach unitarity. Our
results are consistent with Refs. \cite{Efimov_FermiSea_0_0,Efimov_FermiSea_0,Efimov_FermiSea_1,Efimov_FermiSea_2,QCD_Efimov_Cooper},
which explore different, but related scenarios.

\begin{figure*}
\includegraphics[width=0.495\textwidth]{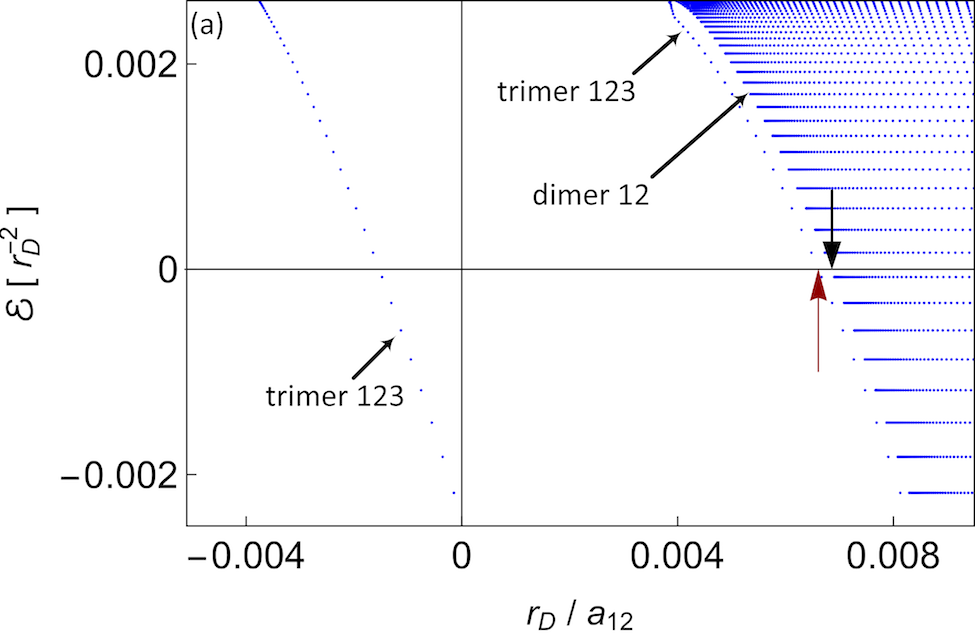}$\quad$\includegraphics[width=0.47\textwidth]{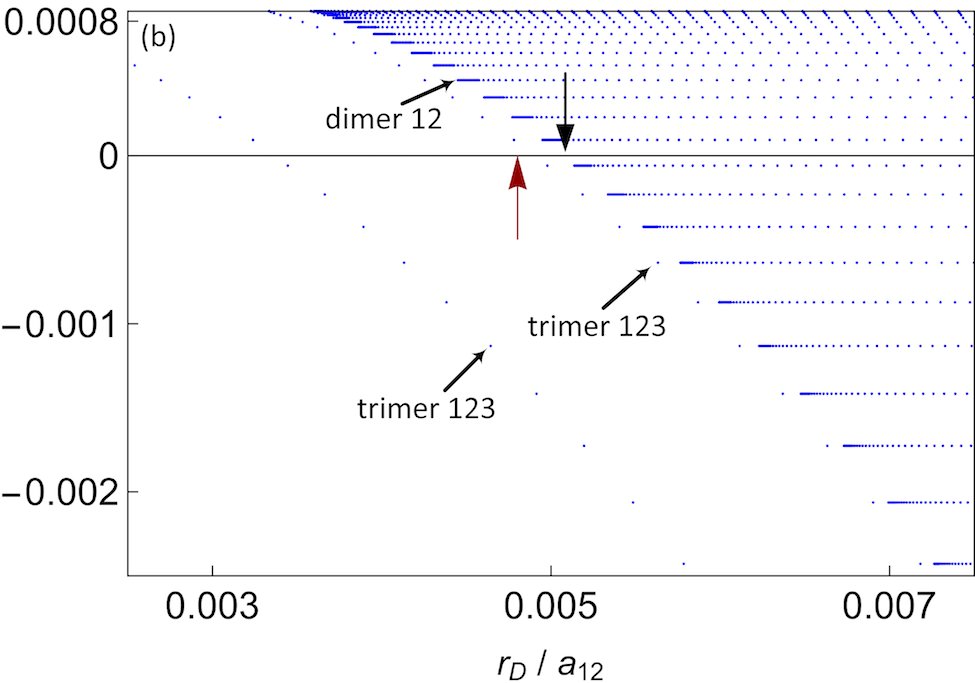}

\caption{A zoom on the plot of energy $\mathcal{E}=2\mu_{12}E/\hbar^{2}$ in
units of $r_{D}^{-2}$ vs $r_{D}/a_{12}$ for (a) $m_{2}/m_{1}\approx6.64$
corresponding to Fig. \ref{Fig4}(a), and (b) $m_{2}/m_{1}\approx22.26$
corresponding to Fig. \ref{Fig4}(b). Both panels show the region
where a highest-energy excited three-body bound state emerges. The
red vertical arrow locates the onset of a highest-energy excited three-body
bound state at zero energy, given by Eq. (\ref{a_12 critical trimer}).
The black vertical arrow locates the onset of the lowest-energy two-body
bound state at zero energy, given by Eq. (\ref{dimer onset at zero energy}). }

\label{Fig5}
\end{figure*}

To find an analytical solution of the lowest-energy two-body bound
state, Cooper pair-23, we note that for $g_{12},g_{13}=0$ the system
of the integral Eqs. (\ref{Main Eq 1}) and (\ref{Main Eq 2}) reduces
to

\begin{equation}
\frac{1}{g_{23}}+\lim_{\Lambda_{2}\rightarrow\infty}\int\frac{d^{3}\mathbf{p}_{3}}{(2\pi)^{3}}\,\frac{\theta_{k_{F},\Lambda_{2}}(\mathbf{p}_{3})}{\frac{\hbar^{2}}{m_{2}}p_{3}^{2}-E_{23}}=0,\label{integral Eq for Cooper pair}
\end{equation}
where $E_{23}<0$ is the bound-state energy of the Cooper pair. We
use the regularization relation (\ref{s wave scattering length})
and solve Eq. (\ref{integral Eq for Cooper pair}) for \emph{s}-wave
symmetry of the states, resulting in

\begin{alignat}{1}
\frac{1}{a_{23}}= & \frac{2}{\pi}k_{F}+\frac{2}{\pi}\sqrt{-\mathcal{E}_{23}}\arctan\left(\frac{\sqrt{-\mathcal{E}_{23}}}{k_{F}}\right),\label{Cooper pair solution}
\end{alignat}
where $\mathcal{E}_{23}=2\mu_{23}E_{23}/\hbar^{2}$ and $\mu_{23}$
is a reduced mass, $1/\mu_{23}=1/m_{2}+1/m_{3}=2/m_{2}$; see gray
dashed curves in Fig. \ref{Fig2}. Far from the resonance, the Cooper-pair
solution for $k_{F}\neq0$ converges asymptotically to the lowest-energy
two-body bound state in vacuum, $1/a_{23}=\sqrt{-\mathcal{E}_{23}}$,
described by Eq. (\ref{Cooper pair solution}) as $k_{F}\rightarrow0$.

For a noninteracting mixture, $g_{23}=0$, Eq. (\ref{Main Eq 2})
has no effect anymore. For \emph{s}-wave symmetry of the states the
integral Eq. (\ref{Main Eq 1}) reduces to

\begin{align}
\Omega_{12}F_{2}(k_{2})= & -\frac{1}{2\pi\frac{\mu_{12}}{m_{1}}k_{2}}\int_{k_{F}}^{\Lambda}d\tilde{p}_{3}\,\tilde{p}_{3}\nonumber \\
 & \times\ln\left(\frac{\tilde{p}_{3}^{2}+\frac{2\mu_{12}}{m_{1}}k_{2}\tilde{p}_{3}+k_{2}^{2}-\mathcal{E}}{\tilde{p}_{3}^{2}-\frac{2\mu_{12}}{m_{1}}k_{2}\tilde{p}_{3}+k_{2}^{2}-\mathcal{E}}\right)F_{2}(\tilde{p}_{3}),\label{main Eq for g23 Zero}
\end{align}
where $\mathcal{E}=2\mu_{12}E/\hbar^{2}$, $E$ is the energy of the
three-body bound state, and $\mu_{12}$ is a reduced mass, $1/\mu_{12}=1/m_{1}+1/m_{2}$;
see Appendix D. The analytical calculation of the function $\Omega_{12}$
is given by Eq. (\ref{Omega_12_result}). We solve the integral Eq.
(\ref{main Eq for g23 Zero}) numerically, using the Gauss-Legendre
quadrature rule; see Appendix E. Figure \ref{Fig3} shows the result
for vanishing and nonvanishing $k_{F}$, where $m_{2}/m_{1}=1$. In
the presence of the Fermi seas, the onset of the three-body bound
state is pushed to positive values of $a_{12}$, and the three-body
bound-state solution converges asymptotically to the corresponding
Efimov state in vacuum. 

\begin{figure*}[t]
\includegraphics[width=1\textwidth]{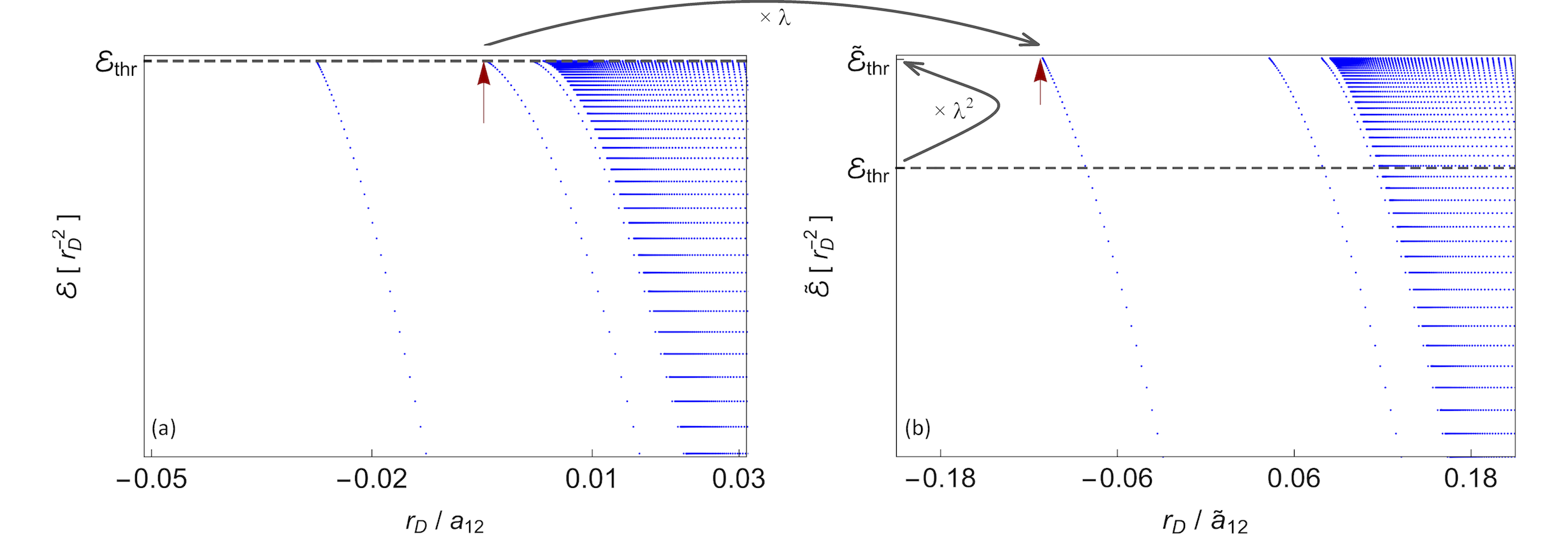}

\caption{Demonstration of the generalized scaling law (\ref{scaling_law_1})
and (\ref{scaling_law_2}) for $g_{23}=0$ and $m_{2}/m_{1}\approx22.26$:
(a) energy $\mathcal{E}=2\mu_{12}E/\hbar^{2}$ in units of $r_{D}^{-2}$
vs $r_{D}/a_{12}$ for $k_{F}r_{D}\approx0.01$, (b) rescaled energy
$\mathcal{\tilde{E}}=2\mu_{12}\tilde{E}/\hbar^{2}$ in units of $r_{D}^{-2}$
vs rescaled $r_{D}/\tilde{a}_{12}$, for $k_{i}\protect\mapsto\lambda k_{i}$,
$i=1,2,3$, $k_{F}\protect\mapsto\lambda k_{F}$, $a_{12}\protect\mapsto\lambda^{-1}a_{12}$,
$E\protect\mapsto\lambda^{2}E$, where $\lambda=\exp(\pi/|s_{0}|)\approx4.84998$.
The red vertical arrow in panel (a) locates the onset of the ($n+1$)-th
excited three-body bound state at $\mathcal{E}_{\mathrm{thr}}=2\mu_{12}E_{\mathrm{thr}}/\hbar^{2}$.
The red vertical arrow in (b) locates the onset of the $n$-th excited
three-body bound state of the rescaled spectrum at $\mathcal{\tilde{E}}_{\mathrm{thr}}=\lambda^{2}\mathcal{E}_{\mathrm{thr}}$.
The gray dashed lines in both panels show the value of \emph{$\mathcal{E}_{\mathrm{thr}}$.}}

\label{Fig6}
\end{figure*}

We note that for a given value of $k_{F}$, as we increase the mass
ratio $m_{2}/m_{1}$, excited three-body bound states appear \cite{excited_trimers}.
Figure \ref{Fig4}(a) shows the result for $m_{2}/m_{1}\approx6.64$,
where two excited additional three-body bound states are visible.
In Fig. \ref{Fig4}(b) we increase the mass ratio to $m_{2}/m_{1}\approx22.26$,
and obtain three excited three-body bound states. The red curves in
Fig. \ref{Fig4}(a) and Fig. \ref{Fig4}(b) show the result in vacuum,
which are the Efimov states. The blue curves show the result in the
presence of Fermi seas. Near unitarity the Fermi seas have a noticeable
influence on the spectrum. Far from the resonance and for low energies,
the effect of the Fermi seas is negligible. In the presence of the
Fermi seas the translational invariance is broken, and the Efimov
scaling law in vacuum does not hold anymore, which we discuss in the
following.

For $g_{23}=0$ we describe the two-body bound-state continuum, dimers-12,
by solving 

\begin{equation}
\Omega_{12}(a_{12},k_{2};k_{F},\mathcal{E}_{12})=0,\label{solution of dimers-12}
\end{equation}
where $\Omega_{12}$ is given by Eq. (\ref{Omega_12_result}), $\mathcal{E}_{12}=2\mu_{12}E_{12}/\hbar^{2}$,
and $E_{12}$ is the energy of the dimers-12. For the lowest-energy
dimer-12 we solve Eq. (\ref{solution of dimers-12}) as $k_{2}\rightarrow k_{F}$.
The result converges asymptotically to the lowest-energy two-body
bound-state solution in vacuum; see gray dashed curves in Fig. \ref{Fig3}
for $m_{2}/m_{1}=1$. At zero energy we find an analytical estimate
for the onset of the the lowest-energy two-body bound state. For that,
we solve Eq. (\ref{solution of dimers-12}) as $k_{2}\rightarrow k_{F}$
and $\mathcal{E}_{12}\rightarrow0$, resulting in a critical \emph{s}-wave
scattering length, $a_{12,\mathrm{dimer}}^{(c)}\equiv a_{12}(E_{12}=0)$:

\begin{alignat}{1}
\frac{1}{a_{12,\mathrm{dimer}}^{(c)}}= & \frac{k_{F}}{\pi}\left[1+\frac{1+\frac{2m_{2}}{m_{1}}}{\frac{2m_{2}}{m_{1}}(1+\frac{m_{2}}{m_{1}})}\ln\left(1+\frac{2m_{2}}{m_{1}}\right)\right.\nonumber \\
 & \left.+\frac{\pi}{2}\frac{1}{1+\frac{m_{2}}{m_{1}}}\sqrt{1+\frac{2m_{2}}{m_{1}}}\right].\label{dimer onset at zero energy}
\end{alignat}
Equation (\ref{dimer onset at zero energy}) gives an estimate of
the shift to the repulsive region of $a_{12}$ that the lowest-energy
two-body bound state undergoes at zero energy in the presence of Fermi
seas; see black vertical arrows in Fig. \ref{Fig5}(a) and \ref{Fig5}(b).
For $m_{2}\gg m_{1}$ this amount approaches $k_{F}/\pi$.

Moreover, for $g_{23}=0$ and a high mass ratio $m_{2}/m_{1}\gg1$,
we find an analytical estimate for the onset of a highest-energy excited
three-body bound state at zero energy. For that, we note that near
the Fermi surface we can approximate the momentum of the species ``2''
and ``3'' to be around $k_{F}$ but in opposite directions, $\mathbf{k}_{2}\sim-\mathbf{k}_{3}$.
Because we have assumed that the total momentum of the three-body
bound state is zero, this results in the vanishing momentum of the
atom ``1'', $\mathbf{k}_{1}\sim\mathbf{0}$. Next, we consider the
pair-12, where $m_{2}/m_{1}\gg1$ and $k_{1}\sim0$. With these assumption,
the relative momentum of the pair-12, defined as $\mathbf{p}_{12}\equiv[m_{2}/(m_{1}+m_{2})]\mathbf{k}_{1}-[m_{1}/(m_{1}+m_{2})]\mathbf{k}_{2}$,
approaches zero. We note that the Fermi surface, $k_{2}\sim k_{F}$,
can be described in terms of the relative momentum, $\mathbf{p}_{12}$,
and total momentum, $\mathbf{P}_{12}$, of the pair-12 as $|(\mu_{12}/m_{1})\mathbf{P}_{12}-\mathbf{p}_{12}|\sim k_{F}$,
where $\mathbf{P}_{12}\equiv\mathbf{k}_{1}+\mathbf{k}_{2}$; see Appendix
F. This implies that for $m_{2}/m_{1}\gg1$ and $k_{1}\sim0$ we can
approximate the total momentum of the pair-12 to be $P_{12}\sim(\mu_{12}/m_{1})^{-1}k_{F}$.
We also note that for large mass ratios $m_{2}/m_{1}$, the threshold
energy of the three-body bound state, $\mathcal{E}_{\mathrm{thr}}=2\mu_{12}E_{\mathrm{thr}}/\hbar^{2}=2(1-\mu_{12}/m_{1})k_{F}^{2}$,
approaches the threshold energy of the pair-12, $\mathcal{E}_{\mathrm{thr}}^{(12)}=\mathcal{E}_{\mathrm{thr}}/2$.
To find the onset of a highest-energy excited three-body bound state
at $E=0$, we calculate the onset of the lowest-energy pair-12 for
total momentum $P_{12}\sim(\mu_{12}/m_{1})^{-1}k_{F}$ and $E_{12}\sim0$.
To do this, we use the interaction model (\ref{Cooper interaction operator}),
and write the Schr{\"o}dinger equation describing the pair-12 for
a contact interaction in terms of the relative and total momenta;
see Appendix F. The solution for $P_{12}\rightarrow(\mu_{12}/m_{1})^{-1}k_{F}$
and $E_{12}\rightarrow0$ results in an estimate for the critical
\emph{s}-wave scattering length, $a_{12,\mathrm{trimer}}^{(c)}\equiv a_{12}(E\approx0)$:

\begin{align}
\frac{1}{a_{12,\mathrm{trimer}}^{(c)}}\approx & \frac{k_{F}}{\pi}\left[1+\frac{1}{4}\frac{1}{1+\frac{m_{2}}{m_{1}}}\ln\left(4(1+\frac{m_{2}}{m_{1}})\right)\right.\nonumber \\
 & \left.-\frac{\pi}{2}\frac{1}{\sqrt{1+\frac{m_{2}}{m_{1}}}}+\frac{1}{2}\frac{1}{1+\frac{m_{2}}{m_{1}}}\right]\text{for }\frac{m_{2}}{m_{1}}\gg1;\label{a_12 critical trimer}
\end{align}
see Appendix F. For a high mass ratio $m_{2}/m_{1}\gg1$, Eq. (\ref{a_12 critical trimer})
gives an estimate for the amount of the shift to the repulsive region
of $a_{12}$ that a highest-energy excited three-body bound state
undergoes at zero energy in the presence of Fermi seas. Figure \ref{Fig5}
reveals a zoom on the region where a highest-energy three-body bound
state emerges for $m_{2}/m_{1}=6.64$ and $m_{2}/m_{1}\approx22.26$.
The red vertical arrows locate the critical value (\ref{a_12 critical trimer}).
For a very large mass ratio $m_{2}/m_{1}$, the critical value (\ref{a_12 critical trimer})
eventually approaches $k_{F}/\pi$, converging to the lowest-energy
two-body bound state at zero energy. Equations (\ref{dimer onset at zero energy})
and (\ref{a_12 critical trimer}) provide a quantitative analysis
for the effect of the Fermi seas on the near-resonant spectrum. 

Finally, we elaborate on the observation that the Fermi seas deform
the Efimov spectrum. This effect is more pronounced as we approach
unitarity. As a result, the Efimov scaling factor that governs the
three-body bound states in vacuum does not hold anymore. Here we show
that a scaling transformation $k_{F}\mapsto\lambda k_{F}$, where
$\lambda$ is the Efimov scaling factor, gives rise to a generalized
scaling law for our system and interaction model (\ref{Cooper interaction operator}).
To this end, we notice that $k_{F}\mapsto\lambda k_{F}$ implies a
scaling transformation of all momenta as $k_{i}\mapsto\lambda k_{i}$,
for $i=1,2,3$. It also rescales the threshold energy as $E_{\mathrm{thr}}\mapsto\lambda^{2}E_{\mathrm{thr}}$,
cf. Eq. (\ref{threshold energy}), implying a general scaling transformation
of energy as $E\mapsto\lambda^{2}E$. To ensure that the system of
the coupled integral Eqs. (\ref{Main Eq 1}) and (\ref{Main Eq 2})
remains valid, it requires a scaling transformation of the \emph{s}-wave
scattering length as $a\mapsto\lambda^{-1}a$; see Eqs. (\ref{Omega_first_case_result}),
(\ref{Omega_second_case_result}), and (\ref{Omega_12_result}). This
results in a discrete scaling law for the three-body bound states
in the presence of Fermi seas:
\begin{alignat}{1}
\frac{\lambda}{a_{n+1}(k_{F})}= & \frac{1}{a_{n}(\lambda k_{F})},\label{scaling_law_1}\\
\lambda^{2}E_{n+1}(k_{F},1/a)= & E_{n}(\lambda k_{F},\lambda/a),\label{scaling_law_2}
\end{alignat}
where $n\in\mathbb{N}$ is an index labeling the three-body bound
state, $\lambda=\exp(\pi/|s_{0}|)$, and the parameter $s_{0}$, that
depends on the mass ratio $m_{2}/m_{1}$, is determined in Appendix
G. Our finding is in agreement with the result of Ref. \cite{Efimov_FermiSea_4}.
Figure \ref{Fig6} demonstrates the generalized scaling law (\ref{scaling_law_1})
and (\ref{scaling_law_2}) for an atomic system of three fermions
with a noninteracting mixture, $g_{23}=0$, and $m_{2}/m_{1}\approx22.26$.

\begin{figure}
\includegraphics[width=1\columnwidth]{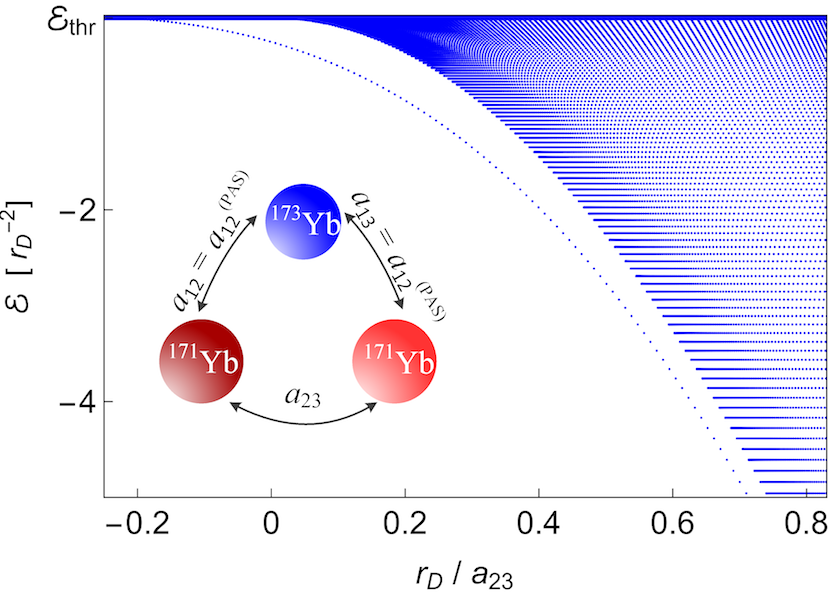}

\caption{Visualization of the first scenario for the experimental signature
of a three-body bound state in an ultracold fermionic mixture of $\text{Yb}$
isotopes. The plot shows the energy $\mathcal{E}=2\mu^{(\mathrm{Yb})}E/\hbar^{2}$
in units of $r_{D}^{-2}$ vs $r_{D}/a_{23}$, where $r_{D}\equiv\ell_{23}^{(\mathrm{vdW})}\approx4.145\text{ nm}$.
The \emph{s}-wave scattering length of $^{171}\mathrm{Yb}$ and $^{173}\mathrm{Yb}$
is fixed as the value measured via photoassociation spectroscopy (PAS),
$a_{12}=a_{13}=a_{12}^{(\mathrm{PAS})}\approx-30.6\text{ nm}$. The
three-body bound state emerges at $a_{23}\approx-20.7\text{ nm}$
at the threshold energy $E_{\mathrm{thr}}\approx1.10\text{ kHz}$.}

\label{Fig7}
\end{figure}

\section{experimental signatures \label{Sec_IV}}

We propose three scenarios to observe three-body bound states in mixtures
of Yb isotopes, in particular a mixture of $^{171}\mathrm{Yb}$ and
$^{173}\mathrm{Yb}$. In the terminology that is illustrated in Fig.
\ref{Fig1}, $^{173}\mathrm{Yb}$ plays the role of species ``1'',
and species ``2'' and ``3'' are two internal states of $^{171}\mathrm{Yb}$.
The density of each of the $^{171}\mathrm{Yb}$ species is $n_{\mathrm{tot}}/2$,
whereas the density of $^{173}\mathrm{Yb}$ is much smaller. We denote
the \emph{s}-wave scattering lengths of $^{171}\mathrm{Yb}$ and $^{173}\mathrm{Yb}$
by $a_{12}$ and $a_{13}$, and the \emph{s}-wave scattering length
of two $^{171}\mathrm{Yb}$ isotopes by $a_{23}$. We also assume
that $a_{13}=a_{12}$.

As measured via two-color photoassociation spectroscopy (PAS), see
Ref. \cite{C6_coeff_Yb_1}, $^{171}\mathrm{Yb}$ isotopes are almost
noninteracting, while the \emph{s}-wave scattering length between
$^{171}\mathrm{Yb}$ and $^{173}\mathrm{Yb}$ atoms is $a_{12}^{(\mathrm{PAS})}\approx-30.6\text{ nm}\approx-578.23a_{0}$,
where $a_{0}$ denotes the Bohr radius \cite{Bohr_radius}. We note
that $^{171}\mathrm{Yb}$ and $^{173}\mathrm{Yb}$ have almost the
same atomic mass, where the reduced mass is $\mu_{12}\approx85.9657\text{ u}$
\cite{NIST_data}. The reduced mass of two $^{171}\mathrm{Yb}$ isotopes
is $\mu_{23}\approx85.4682\text{ u}$ \cite{NIST_data}. The van der
Waals dispersive coefficient, $C_{6}^{(\mathrm{Yb})}$, that determines
the atomic interaction in a $\mathrm{Yb}_{2}$ molecule is given by
Refs. \cite{C6_coeff_Yb_1,C6_coeff_Yb_2}. We calculate the van der
Waals lengths to be $\ell_{12}^{(\mathrm{vdW})}=\frac{1}{2}[2\mu_{12}C_{6}^{(\mathrm{Yb})}/\hbar^{2}]^{1/4}\approx4.151\text{ nm}\approx78.44a_{0}$
and $\ell_{23}^{(\mathrm{vdW})}=\frac{1}{2}[2\mu_{23}C_{6}^{(\mathrm{Yb})}/\hbar^{2}]^{1/4}\approx4.145\text{ nm}\approx78.33a_{0}$.
These values fix the corresponding length scales $r_{D}$. Next, for
each internal state we assume that the density of $^{171}\mathrm{Yb}$
species is $n_{\mathrm{tot}}/2=\frac{1}{2}\times10^{17}\text{ m}^{-3}$.
We calculate the value of the Fermi momentum as $k_{F}=(3\pi^{2}n_{\mathrm{tot}})^{1/3}$;
cf. Ref. \cite{kF_and_n}. 

\begin{figure}
\includegraphics[width=1\columnwidth]{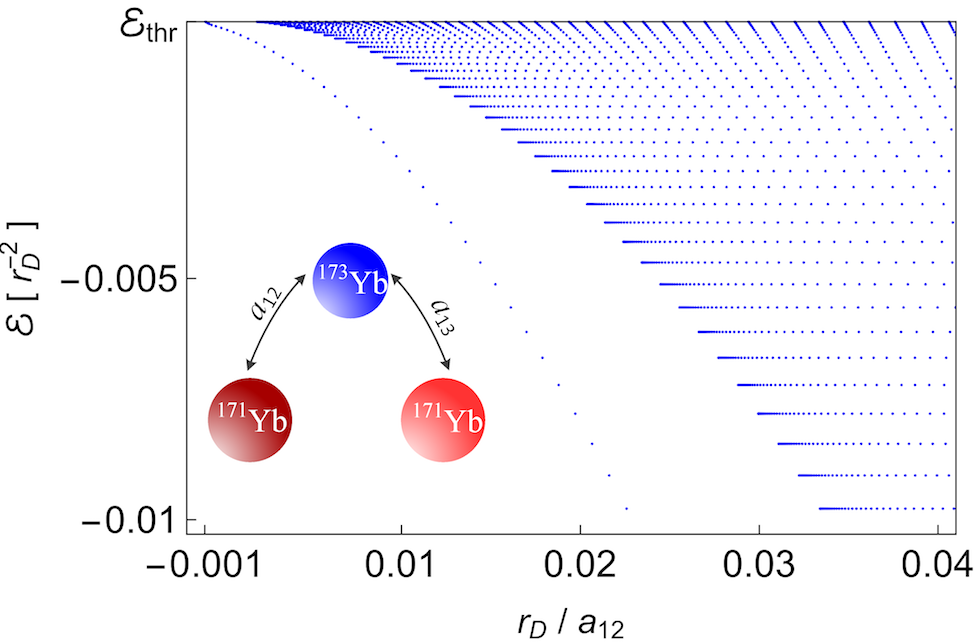}

\caption{Visualization of the second scenario in which $^{171}\mathrm{Yb}$
and $^{173}\mathrm{Yb}$ are interact attractively, while the two
$^{171}\mathrm{Yb}$ species are noninteracting. The plot shows the
energy $\mathcal{E}=2\mu^{(\mathrm{Yb})}E/\hbar^{2}$ in units of
$r_{D}^{-2}$ vs $r_{D}/a_{12}$, where $r_{D}\equiv\ell_{12}^{(\mathrm{vdW})}\approx4.151\text{ nm}$
and $a_{13}=a_{12}$. The onset of the three-body bound state is at
$a_{12}\approx-3193\text{ nm}$ with the threshold energy $E_{\mathrm{thr}}\approx1.09\text{ kHz}$.}

\label{Fig8}
\end{figure}

We adopt the \emph{s}-wave scattering length of $^{171}\mathrm{Yb}$
and $^{173}\mathrm{Yb}$ as reported in Ref. \cite{C6_coeff_Yb_1},
i.e., $a_{12}=a_{12}^{(\mathrm{PAS})}$, and calculate the three-body
bound-state solution for three interacting pairs \cite{Error}. Figure
\ref{Fig7} shows the three-body bound-state energy as a function
of $1/a_{23}$. We find that the onset of the three-body bound state
is $a_{23}\approx-20.7\text{ nm}\approx-391.16a_{0}$, emerging at
the threshold energy $E_{\mathrm{thr}}\approx1.10\text{ kHz}$. As
a first experimental scenario, we propose to tune the interaction
between two $^{171}\mathrm{Yb}$ isotopes via optical Feshbach resonances
\cite{optical_Feshbach_1,optical_Feshbach_2,optical_Feshbach_3,optical_Feshbach_4,optical_Feshbach_5},
across the onset of the three-body bound state, which should result
in increased atomic losses.

As a second scenario we consider two noninteracting $^{171}\mathrm{Yb}$
isotopes, and calculate the three-body bound-state solution for two
interacting pairs $^{171}\mathrm{Yb}$ - $^{173}\mathrm{Yb}$. Figure
\ref{Fig8} shows the energy of the three-body bound state as a function
of $1/a_{12}$. It reveals that the three-body bound state emerges
at $a_{12}\approx-3193\text{ nm}\approx-60336.40a_{0}$ at the threshold
energy $E_{\mathrm{thr}}\approx1.09\text{ kHz}$. Here the \emph{s}-wave
scattering length $a_{12}$ is much larger in amplitude than $a_{12}^{(\mathrm{PAS})}$,
and the threshold energy is smaller than the value obtained in the
first scenario. A three-body bound state is observed, if the interaction
between two $^{171}\mathrm{Yb}$ and $^{173}\mathrm{Yb}$ is tuned
via interisotope Feshbach resonances \cite{interisotope_Feshbach_1},
or via orbital Feshbach resonances \cite{orbital_Feshbach_2,orbital_Feshbach_3}. 

\begin{figure}
\includegraphics[width=1\columnwidth]{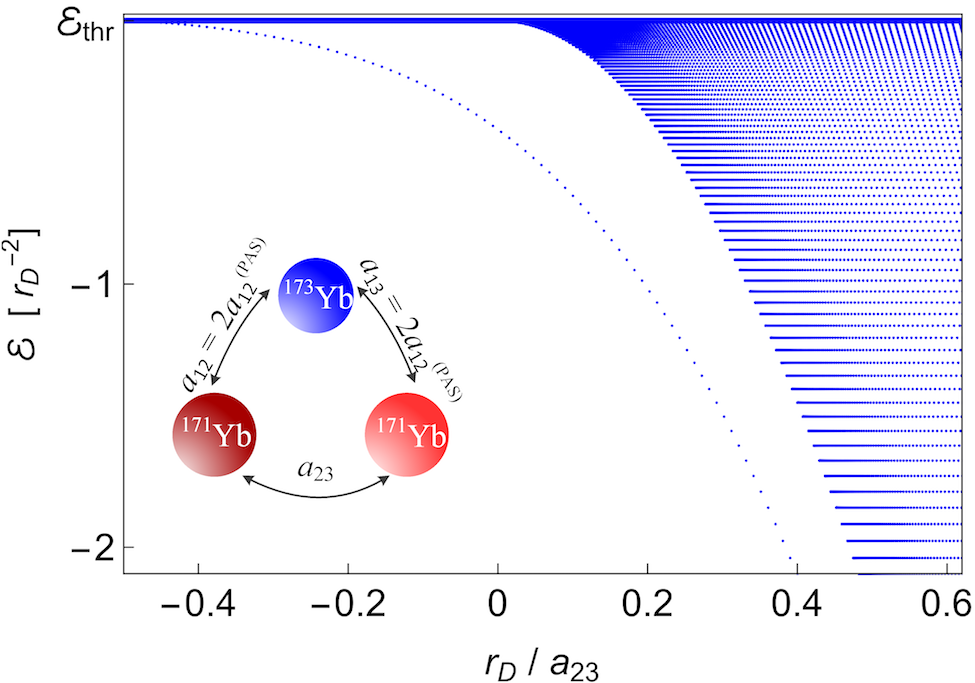}

\caption{Visualization of the third scenario. The plot shows the energy $\mathcal{E}=2\mu^{(\mathrm{Yb})}E/\hbar^{2}$
in units of $r_{D}^{-2}$ vs $r_{D}/a_{23}$, where $r_{D}\equiv\ell_{23}^{(\mathrm{vdW})}\approx4.145\text{ nm}$.
The \emph{s}-wave scattering length of $^{171}\mathrm{Yb}$ and $^{173}\mathrm{Yb}$
is fixed to be $a_{12}=a_{13}=2a_{12}^{(\mathrm{PAS})}\approx-2\times30.6\text{ nm}$.
The three-body bound state emerges at $a_{23}\approx-10.4\text{ nm}$
at the threshold energy $E_{\mathrm{thr}}\approx1.10\text{ kHz}$.}

\label{Fig9}
\end{figure}

As a third scenario, if the interaction between two $^{171}\mathrm{Yb}$
and $^{173}\mathrm{Yb}$ isotopes is tuned to a larger value in amplitude
than $a_{12}^{(\mathrm{PAS})}$, e.g., $a_{12}=2a_{12}^{(\mathrm{PAS})}$,
we find that the three-body bound state emerges at $a_{23}\approx-10.4\text{ nm}\approx-196.52a_{0}$
with the same threshold energy of the first scenario; see Fig. \ref{Fig9}.
Here the value of $a_{12}$ is much smaller in amplitude than the
value obtained in the second scenario. Also, the value of $a_{23}$
is smaller in amplitude than the value found in the first scenario.
A three-body bound state is observed, if the interaction between two
$^{171}\mathrm{Yb}$ isotopes and also the interaction between $^{171}\mathrm{Yb}$
and $^{173}\mathrm{Yb}$ are tuned simultaneously.

We note that in all scenarios we have assumed that the interatomic
distances are much larger than the range of the atomic interactions,
$1/k_{F}\gg r_{D}$. The onset of the three-body bound states might
slightly deviate if this criterion is not met. Here there will be
a competition of $^{171}\mathrm{Yb}$ isotopes to form a three-body
bound state with $^{173}\mathrm{Yb}$.

\section{conclusions\label{Sec_V}}

In conclusion, we have demonstrated and characterized three-body bound
states of a single fermionic atom interacting with a Fermi mixture
of two fermionic species. For this purpose, we have expanded and elaborated
on a model previously used to determine trimer states in conventional
superconductors, Ref. \cite{Sanayei}. We have shown that the expanded
interaction model is separable, leading to a system of integral equations
in momentum space. Based on these equations we have presented their
full numerical solution, as well as analytical solutions of limiting
cases. Compared to three atoms interacting in vacuum, the presence
of the Fermi seas renormalizes the eigenstates and eigenenergies,
in particular near unitarity. Compared the Efimov scaling law of three
atoms in vacuum, we have shown that our system and interaction model
obeys a generalized discrete scaling law. We have also proposed three
scenarios to obtain experimental signatures of the modified Efimov
effect in an ultracold Fermi system of Yb isotopes.

\subsection*{ACKNOWLEDGMENTS }

We acknowledge support from the Deutsche Forschungsgemeinschaft through
Program No. SFB 925, and also The Hamburg Centre for Ultrafast Imaging.
We would like to thank Pascal Naidon for very valuable discussions.
A.S. also thanks C. Becker, K. Sponselee, B. Abeln, and L. Freystatzky
for useful discussions on the experimental signatures of our results. 

\setcounter{equation}{0} \renewcommand{\theequation}{A\arabic{equation}}

\section*{appendix a. introducing the {\large{}{\lowercase{\textit{s}}}}-wave
scattering lengths}

We consider the Schr{\"o}dinger equation in momentum space governing
two atoms ``A'' and ``B'' in vacuum:

\begin{equation}
\left(\frac{\hbar^{2}k_{\mathrm{A}}^{2}}{2m_{\mathrm{A}}}+\frac{\hbar^{2}k_{\mathrm{B}}^{2}}{2m_{\mathrm{B}}}+\hat{U}_{\mathrm{AB}}-E_{\mathrm{AB}}\right)\phi=0,\label{Schrodinger Eq for atoms a and b}
\end{equation}
where $m_{i}$ and $\mathbf{k}_{i}$, $i\in\{\mathrm{A},\mathrm{B}\}$,
is the atom mass and momentum, respectively, $E_{\mathrm{AB}}$ is
the energy, and $\phi=\phi(\mathbf{k}_{\mathrm{A}},\mathbf{k}_{\mathrm{B}})$
is the wave function. The interaction $\hat{U}_{\mathrm{AB}}$ between
the atoms ``A'' and ``B'' follows from the interaction model (\ref{Cooper interaction operator}).
The resulting operator $\hat{U}_{\mathrm{AB}}\phi$ reads:

\begin{align}
\hat{U}_{\mathrm{AB}}\phi= & g_{\mathrm{AB}}\theta_{\Lambda_{\mathrm{A}}}(\mathbf{k}_{\mathrm{A}})\theta_{\Lambda_{\mathrm{B}}}(\mathbf{k}_{\mathrm{B}})\int\frac{d^{3}\mathbf{q}}{(2\pi)^{3}}\,\theta_{\Lambda_{\mathrm{A}}}(\mathbf{k}_{\mathrm{A}}-\mathbf{q})\nonumber \\
 & \times\theta_{\Lambda_{\mathrm{B}}}(\mathbf{k}_{\mathrm{B}}+\mathbf{q})\phi(\mathbf{k}_{\mathrm{A}}-\mathbf{q},\mathbf{k}_{\mathrm{B}}+\mathbf{q}),\label{V_ab}
\end{align}
where $g_{\mathrm{AB}}<0$ and $\mathbf{q}$ is the momentum transfer
\cite{momentum_transfer}. We assume the zero total momentum, $\mathbf{k}_{\mathrm{A}}+\mathbf{k}_{\mathrm{B}}=\mathbf{0}$,
and $\Lambda_{\mathrm{B}}=\Lambda_{\mathrm{A}}$. Next, we define
the variables $\boldsymbol{\kappa}_{i}\equiv\mathbf{q}+\mathbf{k}_{i}$,
$i\in\{\mathrm{A},\mathrm{B}\}$, and write the Schr{\"o}dinger Eq.
(\ref{Schrodinger Eq for atoms a and b}) as

\begin{align}
\left(\frac{\hbar^{2}k_{\mathrm{A}}^{2}}{2\mu_{\mathrm{AB}}}-E_{\mathrm{AB}}\right)\phi(\mathbf{k}_{\mathrm{A}})= & -g_{\mathrm{AB}}\theta_{\Lambda_{\mathrm{A}}}(\mathbf{k}_{\mathrm{A}})\int\frac{d^{3}\boldsymbol{\kappa}_{\mathrm{B}}}{(2\pi)^{3}}\nonumber \\
 & \times\theta_{\Lambda_{\mathrm{A}}}(\boldsymbol{\kappa}_{\mathrm{B}})\phi(\boldsymbol{\kappa}_{\mathrm{B}}),\label{Schrodinger Eq for electrons a and b, V2}
\end{align}
where $\mu_{\mathrm{AB}}$ is a reduced mass, $1/\mu_{\mathrm{AB}}=1/m_{\mathrm{A}}+1/m_{\mathrm{B}}$.
We define 

\begin{equation}
\mathcal{F}\equiv-4\pi\left(\frac{2\mu_{\mathrm{AB}}}{4\pi\hbar^{2}}g_{\mathrm{AB}}\right)\theta_{\Lambda_{\mathrm{A}}}(\mathbf{k}_{\mathrm{A}})\int\frac{d^{3}\mathbf{k}_{\mathrm{A}}}{(2\pi)^{3}}\theta_{\Lambda_{\mathrm{A}}}(\mathbf{k}_{\mathrm{A}})\phi(\mathbf{k}_{\mathrm{A}}),\label{function F}
\end{equation}
and rewrite Eq. (\ref{Schrodinger Eq for electrons a and b, V2})
as

\begin{equation}
(k_{\mathrm{A}}^{2}-\mathcal{E}_{\mathrm{AB}})\phi(\mathbf{k}_{\mathrm{A}})=\mathcal{F},\label{Schrodinger Eq for electrons a and b, V3}
\end{equation}
where $\mathcal{E}_{\mathrm{AB}}=2\mu_{\mathrm{AB}}E_{\mathrm{AB}}/\hbar^{2}$.

For $\mathcal{E}_{\mathrm{AB}}>0$ the solution of Eq. (\ref{Schrodinger Eq for electrons a and b, V3})
is

\begin{equation}
\phi(\mathbf{k}_{\mathrm{A}})=(2\pi)^{3}\delta^{(3)}(\mathbf{k}_{\mathrm{A}}-\mathbf{K})+\frac{\mathcal{F}}{k_{\mathrm{A}}^{2}-\mathcal{E}_{\mathrm{AB}}+i\varepsilon},\label{ansatz for wave function E>0}
\end{equation}
where $0<\varepsilon\ll1$, $|\mathbf{K}|^{2}=\mathcal{E}_{\mathrm{AB}}$,
and $\delta^{(3)}$ denotes the three-dimensional Dirac delta function.
We insert the ansazt (\ref{ansatz for wave function E>0}) into Eq.
(\ref{function F}):

\begin{alignat}{1}
\frac{\mathcal{F}}{4\pi\left(\frac{2\mu_{\mathrm{AB}}}{4\pi\hbar^{2}}g_{\mathrm{AB}}\right)}= & -\theta_{\Lambda_{\mathrm{A}}}(\mathbf{k}_{\mathrm{A}})\int\frac{d^{3}\mathbf{k}_{\mathrm{A}}}{(2\pi)^{3}}\theta_{\Lambda_{\mathrm{A}}}(\mathbf{k}_{\mathrm{A}})\Bigl[(2\pi)^{3}\nonumber \\
 & \times\delta^{(3)}(\mathbf{k}_{\mathrm{A}}-\mathbf{K})+\frac{\mathcal{F}}{k_{\mathrm{A}}^{2}-\mathcal{E}_{\mathrm{AB}}+i\varepsilon}\Bigr].\label{ansatz in F}
\end{alignat}
We note that in the zero-energy limit, $\mathcal{E}_{\mathrm{AB}}\rightarrow0^{+}$,
we have $\mathcal{F}=-4\pi a_{\mathrm{AB}}$, where $a_{\mathrm{AB}}$
is the \emph{s}-wave scattering length; see Ref. \cite{Scattering_Book}.
Next, for contact interactions and \emph{s}-wave symmetry of the states,
we evaluate Eq. (\ref{ansatz in F}) by taking the limit of $\Lambda_{\mathrm{A}}$
to infinity:

\begin{equation}
\frac{4\pi\hbar^{2}}{2\mu_{\mathrm{AB}}g_{\mathrm{AB}}}+\frac{2}{\pi}\lim_{\Lambda_{\mathrm{A}}\rightarrow\infty}\int_{0}^{\Lambda_{\mathrm{A}}}dk_{\mathrm{A}}\,\frac{k_{\mathrm{A}}^{2}}{k_{\mathrm{A}}^{2}+i\varepsilon}=\frac{1}{a_{\mathrm{AB}}},\label{equation for a and g}
\end{equation}
which yields

\begin{equation}
\frac{2\pi\hbar^{2}}{\mu_{\mathrm{AB}}}\frac{1}{g_{\mathrm{AB}}}+\frac{2}{\pi}\Lambda_{\mathrm{A}}=\frac{1}{a_{\mathrm{AB}}}\text{ as }\Lambda_{\mathrm{A}}\rightarrow\infty.\label{renormalization Eq}
\end{equation}
In this paper, we use Eq. (\ref{renormalization Eq}) as a regularization
relation to introduce the \emph{s}-wave scattering length. With this,
we can eliminate the ultraviolet divergences due to contact interactions.

We also notice that for the bound states, $\mathcal{E}_{\mathrm{AB}}<0$,
the solution of Eq. (\ref{Schrodinger Eq for electrons a and b, V3})
is

\begin{equation}
\phi(\mathbf{k}_{\mathrm{A}})=\frac{\mathcal{F}}{k_{\mathrm{A}}^{2}-\mathcal{E}_{\mathrm{AB}}}.\label{ansatz for wave function E<0}
\end{equation}
We insert the ansatz (\ref{ansatz for wave function E<0}) into Eq.
(\ref{function F}), take the limit $\Lambda_{\mathrm{A}}\rightarrow\infty$,
and use Eq. (\ref{renormalization Eq}). This results in

\begin{equation}
\frac{1}{a_{\mathrm{AB}}}=\sqrt{-\mathcal{E}_{\mathrm{AB}}};\label{first dimer in vacuum}
\end{equation}
cf. Fig. \ref{Fig10}. Equation (\ref{first dimer in vacuum}) shows
that for contact interactions the lowest-energy two-body bound state
in vacuum emerges at unitarity, $a_{\mathrm{AB}}\rightarrow\pm\infty$,
where $|\mathcal{E}_{\mathrm{AB}}|\rightarrow0^{+}$; cf. Figs. \ref{Fig2}
and \ref{Fig3}. 

\setcounter{equation}{0} \renewcommand{\theequation}{B\arabic{equation}}

\section*{appendix b. separable interaction model (\ref{Cooper interaction operator})
and derivation of the system of two coupled integral eqs. (\ref{Main Eq 1})
and (\ref{Main Eq 2})}

We apply the interaction operators $\hat{U}_{ij}$, given by Eq. (\ref{Cooper interaction operator}),
on the wave function $\psi=\psi(\mathbf{k}_{1},\mathbf{k}_{2},\mathbf{k}_{3})$,
and write the Schr{\"o}dinger Eq. (\ref{main Schrodinger Eq}) as
follows:

\begin{equation}
\left(\frac{\hbar^{2}k_{1}^{2}}{2m_{1}}+\frac{\hbar^{2}k_{2}^{2}}{2m_{2}}+\frac{\hbar^{2}k_{3}^{2}}{2m_{3}}-E\right)\psi=-(\hat{U}_{12}+\hat{U}_{13}+\hat{U}_{23})\psi,\label{integral Eq 1}
\end{equation}
where

\begin{align}
\hat{U}_{12}\psi= & g_{12}\theta_{\Lambda_{1}}(\mathbf{k}_{1})\theta_{\Lambda_{2}}(\mathbf{k}_{2})\int\frac{d^{3}\mathbf{q}}{(2\pi)^{3}}\theta_{\Lambda_{1}}(\mathbf{k}_{1}-\mathbf{q})\nonumber \\
 & \times\theta_{\Lambda_{2}}(\mathbf{k}_{2}+\mathbf{q})\psi(\mathbf{k}_{1}-\mathbf{q},\mathbf{k}_{2}+\mathbf{q},\mathbf{k}_{3}),\label{U12 on psi}
\end{align}

\begin{align}
\hat{U}_{13}\psi= & g_{13}\theta_{\Lambda_{1}}(\mathbf{k}_{1})\theta_{\Lambda_{3}}(\mathbf{k}_{3})\int\frac{d^{3}\mathbf{q}}{(2\pi)^{3}}\theta_{\Lambda_{1}}(\mathbf{k}_{1}-\mathbf{q})\nonumber \\
 & \times\theta_{\Lambda_{3}}(\mathbf{k}_{3}+\mathbf{q})\psi(\mathbf{k}_{1}-\mathbf{q},\mathbf{k}_{2},\mathbf{k}_{3}+\mathbf{q}),\label{U13 on psi}
\end{align}

\begin{align}
\hat{U}_{23}\psi= & g_{23}\theta_{\Lambda_{2}}(\mathbf{k}_{2})\theta_{\Lambda_{3}}(\mathbf{k}_{3})\int\frac{d^{3}\mathbf{q}}{(2\pi)^{3}}\theta_{\Lambda_{2}}(\mathbf{k}_{2}-\mathbf{q})\nonumber \\
 & \times\theta_{\Lambda_{3}}(\mathbf{k}_{3}+\mathbf{q})\psi(\mathbf{k}_{1},\mathbf{k}_{2}-\mathbf{q},\mathbf{k}_{3}+\mathbf{q}),\label{U23 on psi}
\end{align}
and the cutoff function $\theta$ is defined by Eq. (\ref{cutoff func}).
The resulting operators (\ref{U12 on psi})-(\ref{U23 on psi}) reveal
that the interaction operator $\hat{U}$ is separable \cite{separable}.
Next, we define the variables $\boldsymbol{\kappa}_{i}\equiv\mathbf{q}+\mathbf{k}_{i}$,
for $i=1,2,3$, and also assume $m_{3}=m_{2}$ and $\Lambda_{1}\sim\Lambda_{2}=\Lambda_{3}$.
We consider the zero total momentum of the three-body bound states,
$\psi(\mathbf{k}_{1},\mathbf{k}_{2},\mathbf{k}_{3})=\psi(\mathbf{k}_{2},\mathbf{k}_{3})\delta^{(3)}(\mathbf{k}_{1}+\mathbf{k}_{2}+\mathbf{k}_{3})$,
where $\delta^{(3)}$ denotes the three-dimensional Dirac delta function.
We also define three functions $F_{1}$, $F_{2}$, and $F_{3}$ as

\begin{align}
F_{1}(\mathbf{k}_{1})= & g_{23}\int\frac{d^{3}\boldsymbol{\kappa}_{3}}{(2\pi)^{3}}\theta_{\Lambda_{2}}(-\mathbf{k}_{1}-\boldsymbol{\kappa}_{3})\theta_{\Lambda_{3}}(\boldsymbol{\kappa}_{3})\nonumber \\
 & \times\psi(-\mathbf{k}_{1}-\boldsymbol{\kappa}_{3},\boldsymbol{\kappa}_{3}),\label{F1 - V1}\\
F_{2}(\mathbf{k}_{2})= & g_{13}\int\frac{d^{3}\boldsymbol{\kappa}_{3}}{(2\pi)^{3}}\theta_{\Lambda_{1}}(-\mathbf{k}_{2}-\boldsymbol{\kappa}_{3})\theta_{\Lambda_{3}}(\boldsymbol{\kappa}_{3})\psi(\mathbf{k}_{2},\boldsymbol{\kappa}_{3}),\label{F2 - V1}\\
F_{3}(\mathbf{k}_{3})= & g_{12}\int\frac{d^{3}\boldsymbol{\kappa}_{2}}{(2\pi)^{3}}\theta_{\Lambda_{1}}(-\mathbf{k}_{3}-\boldsymbol{\kappa}_{2})\theta_{\Lambda_{2}}(\boldsymbol{\kappa}_{2})\psi(\boldsymbol{\kappa}_{2},\mathbf{k}_{3}).\label{F3 - V1}
\end{align}
We use Eqs. (\ref{F1 - V1})-(\ref{F3 - V1}) and rewrite Eq. (\ref{integral Eq 1})
as follows:

\begin{align}
\left(\frac{\hbar^{2}(\mathbf{k}_{2}+\mathbf{k}_{3})^{2}}{2m_{1}}+\frac{\hbar^{2}k_{2}^{2}}{2m_{2}}+\frac{\hbar^{2}k_{3}^{2}}{2m_{3}}-E\right)\psi(\mathbf{k}_{2},\mathbf{k}_{3})\qquad\qquad\nonumber \\
=-\theta_{\Lambda_{2}}(\mathbf{k}_{2})\theta_{\Lambda_{2}}(\mathbf{k}_{3})F_{1}(-\mathbf{k}_{2}-\mathbf{k}_{3})-\theta_{\Lambda_{1}}(-\mathbf{k}_{2}-\mathbf{k}_{3})\nonumber \\
\times\theta_{\Lambda_{3}}(\mathbf{k}_{3})F_{2}(\mathbf{k}_{2})-\theta_{\Lambda_{1}}(-\mathbf{k}_{2}-\mathbf{k}_{3})\theta_{\Lambda_{2}}(\mathbf{k}_{2})F_{3}(\mathbf{k}_{3}).\label{integral Eq 2}
\end{align}
Equation (\ref{integral Eq 2}) provides an ansatz for the wave function:

\begin{widetext}

\begin{equation}
\psi(\mathbf{k}_{2},\mathbf{k}_{3})=-\frac{\theta_{\Lambda_{2}}(\mathbf{k}_{2})\theta_{\Lambda_{2}}(\mathbf{k}_{3})F_{1}(-\mathbf{k}_{2}-\mathbf{k}_{3})+\theta_{\Lambda_{1}}(-\mathbf{k}_{2}-\mathbf{k}_{3})\theta_{\Lambda_{3}}(\mathbf{k}_{3})F_{2}(\mathbf{k}_{2})+\theta_{\Lambda_{1}}(-\mathbf{k}_{2}-\mathbf{k}_{3})\theta_{\Lambda_{2}}(\mathbf{k}_{2})F_{3}(\mathbf{k}_{3})}{\frac{\hbar^{2}(\mathbf{k}_{2}+\mathbf{k}_{3})^{2}}{2m_{1}}+\frac{\hbar^{2}k_{2}^{2}}{2m_{2}}+\frac{\hbar^{2}k_{3}^{2}}{2m_{3}}-E}.\label{wavefunction ansatz}
\end{equation}

\end{widetext}We take into account the Fermi sea constraints by $k_{2}>k_{F}$
and $k_{3}>k_{F}$. We also assume $g_{13}=g_{12}$. If the species
``2'' and ``3'' are in a singlet state, then $F_{3}=F_{2}$. Now
we define $\mathbf{p}_{1}\equiv-\mathbf{k}_{2}-\boldsymbol{\kappa}_{3}$,
$\mathbf{p}_{2}\equiv-\mathbf{k}_{1}-\boldsymbol{\kappa}_{3}$, $\mathbf{p}_{3}\equiv\boldsymbol{\kappa}_{3}$,
and rewrite the unknown functions $F_{1}$ and $F_{2}$ as follows:

\begin{align}
F_{1}(\mathbf{k}_{1})= & g_{23}\int\frac{d^{3}\mathbf{p}_{3}}{(2\pi)^{3}}\theta_{k_{F},\Lambda_{2}}(-\mathbf{k}_{1}-\mathbf{p}_{3})\theta_{k_{F},\Lambda_{2}}(\mathbf{p}_{3})\nonumber \\
 & \times\psi(-\mathbf{k}_{1}-\mathbf{p}_{3},\mathbf{p}_{3}),\label{F1 _main_representation}\\
F_{2}(\mathbf{k}_{2})= & g_{12}\int\frac{d^{3}\mathbf{p}_{3}}{(2\pi)^{3}}\theta_{\Lambda_{1}}(-\mathbf{k}_{2}-\mathbf{p}_{3})\theta_{k_{F},\Lambda_{2}}(\mathbf{p}_{3})\psi(\mathbf{k}_{2},\mathbf{p}_{3}).\label{F2_main_representation}
\end{align}
Finally, we choose a three-body parameter $\Lambda\gg k_{F}$ to fix
the range of the interactions and to regularize the three-body bound-state
solutions. We insert the ansatz (\ref{wavefunction ansatz}) into
Eqs. (\ref{F1 _main_representation}) and (\ref{F2_main_representation}),
and arrive at the system of two coupled integral Eqs. (\ref{Main Eq 1})
and (\ref{Main Eq 2}), where the integral kernels $K_{i}$ and $\tilde{K_{i}}$,
$i=1,2,3$, are:

\begin{align}
K_{1}(\mathbf{k}_{2},\mathbf{p}_{3};E)= & \frac{\theta_{\Lambda_{1}}(-\mathbf{k}_{2}-\mathbf{p}_{3})\theta_{k_{F},\Lambda_{2}}(\mathbf{p}_{3})}{\frac{\hbar^{2}(\mathbf{k}_{2}+\mathbf{p}_{3})^{2}}{2m_{1}}+\frac{\hbar^{2}k_{2}^{2}}{2m_{2}}+\frac{\hbar^{2}p_{3}^{2}}{2m_{2}}-E},\label{K1}\\
K_{2}(\mathbf{k}_{2},\mathbf{p}_{1};E)= & \frac{\theta_{\Lambda_{1}}(\mathbf{p}_{1})\theta_{k_{F},\Lambda_{2}}(-\mathbf{p}_{1}-\mathbf{k}_{2})}{\frac{\hbar^{2}p_{1}^{2}}{2m_{1}}+\frac{\hbar^{2}k_{2}^{2}}{2m_{2}}+\frac{\hbar^{2}(\mathbf{p}_{1}+\mathbf{k}_{2})^{2}}{2m_{2}}-E},\label{K2}\\
K_{3}(\mathbf{k}_{1},\mathbf{p}_{3};E)= & \frac{\theta_{k_{F},\Lambda_{2}}(-\mathbf{k}_{1}-\mathbf{p}_{3})\theta_{k_{F},\Lambda_{2}}(\mathbf{p}_{3})}{\frac{\hbar^{2}k_{1}^{2}}{2m_{1}}+\frac{\hbar^{2}(\mathbf{k}_{1}+\mathbf{p}_{3})^{2}}{2m_{2}}+\frac{\hbar^{2}p_{3}^{2}}{2m_{2}}-E},\label{K3}
\end{align}

\begin{align}
\tilde{K}_{1}(\mathbf{k}_{2},\tilde{\mathbf{p}}_{3};E)= & \frac{\theta_{\Lambda}(-\mathbf{k}_{2}-\tilde{\mathbf{p}}_{3})\theta_{k_{F},\Lambda}(\mathbf{k}_{2})\theta_{k_{F},\Lambda}(\tilde{\mathbf{p}}_{3})}{\frac{\hbar^{2}(\mathbf{k}_{2}+\tilde{\mathbf{p}}_{3})^{2}}{2m_{1}}+\frac{\hbar^{2}k_{2}^{2}}{2m_{2}}+\frac{\hbar^{2}\tilde{p}_{3}^{2}}{2m_{2}}-E},\label{K1_tilde}\\
\tilde{K}_{2}(\mathbf{k}_{2},\tilde{\mathbf{p}}_{1};E)= & \frac{\theta_{\Lambda}(\tilde{\mathbf{p}}_{1})\theta_{k_{F},\Lambda}(\mathbf{k}_{2})\theta_{k_{F},\Lambda}(-\tilde{\mathbf{p}}_{1}-\mathbf{k}_{2})}{\frac{\hbar^{2}\tilde{p}_{1}^{2}}{2m_{1}}+\frac{\hbar^{2}k_{2}^{2}}{2m_{2}}+\frac{\hbar^{2}(\tilde{\mathbf{p}}_{1}+\mathbf{k}_{2})^{2}}{2m_{2}}-E},\label{K2_tilde}\\
\tilde{K}_{3}(\mathbf{k}_{1},\tilde{\mathbf{p}}_{3};E)= & \frac{\theta_{k_{F},\Lambda}(-\mathbf{k}_{1}-\tilde{\mathbf{p}}_{3})\theta_{\Lambda}(\mathbf{k}_{1})\theta_{k_{F},\Lambda}(\tilde{\mathbf{p}}_{3})}{\frac{\hbar^{2}k_{1}^{2}}{2m_{1}}+\frac{\hbar^{2}(\mathbf{k}_{1}+\tilde{\mathbf{p}}_{3})^{2}}{2m_{2}}+\frac{\hbar^{2}\tilde{p}_{3}^{2}}{2m_{2}}-E}.\label{K3_tilde}
\end{align}

\setcounter{equation}{0} \renewcommand{\theequation}{C\arabic{equation}}

\section*{appendix c. calculation of the function {\large{}$\Omega_{23}$} }

For \emph{s}-wave symmetry of the states we write the integral kernel
$K_{3}(\mathbf{k}_{1},\mathbf{p}_{3};\mathcal{E})$ as

\begin{equation}
\mathcal{K}_{3}(k_{1},p_{3};\mathcal{E})=p_{3}\ln\left(\frac{p_{3}^{2}+k_{1}p_{3}v_{\mathrm{max}}+\frac{\mu_{23}}{\mu_{12}}k_{1}^{2}-\frac{\mu_{23}}{\mu_{12}}\mathcal{E}}{p_{3}^{2}+k_{1}p_{3}v_{\mathrm{min}}+\frac{\mu_{23}}{\mu_{12}}k_{1}^{2}-\frac{\mu_{23}}{\mu_{12}}\mathcal{E}}\right),\label{K3 - V2}
\end{equation}
where $\mathcal{E}=2\mu_{12}E/\hbar^{2}$, $E$ is the energy of the
three-body system, $v_{\mathrm{max}}$ and $v_{\mathrm{min}}$ denote
the upper- and lower bound of $v\equiv\cos\vartheta_{\mathbf{p}_{3},\mathbf{k}_{1}}$,
respectively, and $\mu_{23}$ is a reduced mass, $1/\mu_{23}=1/m_{2}+1/m_{3}=2/m_{2}$.
For contact interactions we have:

\begin{equation}
v_{\mathrm{max}}=\min_{p_{3}}\left(1,\frac{\Lambda_{2}^{2}-k_{1}^{2}-p_{3}^{2}}{2k_{1}p_{3}}\right)\rightarrow1\;\text{as }\Lambda_{2}\rightarrow\infty,\label{v_max}
\end{equation}

\begin{align}
v_{\mathrm{min}}= & \max_{p_{3}}\left(-1,\frac{k_{F}^{2}-k_{1}^{2}-p_{3}^{2}}{2k_{1}p_{3}}\right)\nonumber \\
= & \begin{cases}
-1, & \begin{array}{c}
\text{for }k_{F}<p_{3}<k_{1}-k_{F}\\
\text{ or }p_{3}>k_{1}+k_{F},
\end{array}\\
\\
\frac{k_{F}^{2}-k_{1}^{2}-p_{3}^{2}}{2k_{1}p_{3}}, & \text{for }k_{1}-k_{F}<p_{3}<k_{1}+k_{F}.
\end{cases}\label{v_min}
\end{align}
Next, without loss of generality we assume that $\mathbf{p}_{3}=p_{3}\mathbf{e}_{z}$,
where $\mathbf{e}_{z}$ is the unit vector in the direction of the
$z$-axis, and calculate the function $\Omega_{23}$ for contact interactions:

\begin{alignat}{1}
\Omega_{23}\equiv & \Omega_{23}(a_{23},k_{1};k_{F},\mathcal{E})\nonumber \\
\equiv & \frac{4\pi\hbar^{2}}{2\mu_{23}g_{23}}+\frac{1}{\frac{2\mu_{23}}{m_{2}}\pi k_{1}}\lim_{\Lambda_{2}\rightarrow\infty}\int_{k_{F}}^{\Lambda_{2}}dp_{3}\nonumber \\
 & \times p_{3}\ln\left(\frac{p_{3}^{2}+k_{1}p_{3}v_{\mathrm{max}}+\frac{\mu_{23}}{\mu_{12}}k_{1}^{2}-\frac{\mu_{23}}{\mu_{12}}\mathcal{E}}{p_{3}^{2}+k_{1}p_{3}v_{\mathrm{min}}+\frac{\mu_{23}}{\mu_{12}}k_{1}^{2}-\frac{\mu_{23}}{\mu_{12}}\mathcal{E}}\right).\label{Omega_23_V1}
\end{alignat}
To calculate Eq. (\ref{Omega_23_V1}) we consider two cases. For $0<k_{1}\leqslant2k_{F}$
we have:

\begin{align}
\Omega_{23}= & \frac{4\pi\hbar^{2}}{2\mu_{23}g_{23}}+\frac{1}{\pi k_{1}}\int_{k_{F}}^{k_{1}+k_{F}}dp_{3}\,p_{3}\nonumber \\
 & \qquad\times\ln\left(\frac{p_{3}^{2}+k_{1}p_{3}+\frac{\mu_{23}}{\mu_{12}}k_{1}^{2}-\frac{\mu_{23}}{\mu_{12}}\mathcal{E}}{\frac{1}{2}p_{3}^{2}+(\frac{\mu_{23}}{\mu_{12}}-\frac{1}{2})k_{1}^{2}+\frac{1}{2}k_{F}^{2}-\frac{\mu_{23}}{\mu_{12}}\mathcal{E}}\right)\nonumber \\
 & +\frac{1}{\pi k_{1}}\lim_{\Lambda_{2}\rightarrow\infty}\int_{k_{1}+k_{F}}^{\Lambda_{2}}dp_{3}\,p_{3}\nonumber \\
 & \qquad\times\ln\left(\frac{p_{3}^{2}+k_{1}p_{3}+\frac{\mu_{23}}{\mu_{12}}k_{1}^{2}-\frac{\mu_{23}}{\mu_{12}}\mathcal{E}}{p_{3}^{2}-k_{1}p_{3}+\frac{\mu_{23}}{\mu_{12}}k_{1}^{2}-\frac{\mu_{23}}{\mu_{12}}\mathcal{E}}\right).\label{Omega_first_case_1}
\end{align}
We calculate each integral and use Eq. (\ref{s wave scattering length}).
The result is

\begin{align}
\Omega_{23}= & \frac{1}{a_{23}}-\frac{k_{1}}{2\pi}-\frac{k_{F}}{\pi}+\frac{2\sqrt{\kappa}}{\pi}\left[\arctan\left(\frac{\frac{1}{2}k_{1}+k_{F}}{\sqrt{\kappa}}\right)-\frac{\pi}{2}\right]\nonumber \\
 & +\frac{1}{\pi k_{1}}\left((\frac{\mu_{23}}{\mu_{12}}-\frac{1}{2})k_{1}^{2}+k_{F}^{2}-\frac{\mu_{23}}{\mu_{12}}\mathcal{E}\right)\nonumber \\
 & \times\ln\left(\frac{(\frac{\mu_{23}}{\mu_{12}}-\frac{1}{2})k_{1}^{2}+k_{F}^{2}-\frac{\mu_{23}}{\mu_{12}}\mathcal{E}}{\frac{\mu_{23}}{\mu_{12}}k_{1}^{2}+k_{F}k_{1}+k_{F}^{2}-\frac{\mu_{23}}{\mu_{12}}\mathcal{E}}\right),\label{Omega_first_case_result}
\end{align}
where $\kappa\equiv(\frac{\mu_{23}}{\mu_{12}}-\frac{1}{4})k_{1}^{2}-\frac{\mu_{23}}{\mu_{12}}\mathcal{E}$.
The lowest-energy two-body bound state, Cooper pair-23, is described
by

\begin{equation}
\Omega_{23}(a_{23},k_{1}\rightarrow0;k_{F},\mathcal{E}\rightarrow\mathcal{E}_{23})=0,\label{first dimer23}
\end{equation}
resulting in Eq. (\ref{Cooper pair solution}); cf. Fig. \ref{Fig2}.

For $k_{1}\geqslant2k_{F}$ we have:

\begin{align}
\Omega_{23}= & \frac{4\pi\hbar^{2}}{2\mu_{23}g_{23}}+\frac{1}{\pi k_{1}}\int_{k_{F}}^{k_{1}-k_{F}}dp_{3}\,p_{3}\nonumber \\
 & \qquad\times\ln\left(\frac{p_{3}^{2}+k_{1}p_{3}+\frac{\mu_{23}}{\mu_{12}}k_{1}^{2}-\frac{\mu_{23}}{\mu_{12}}\mathcal{E}}{p_{3}^{2}-k_{1}p_{3}+\frac{\mu_{23}}{\mu_{12}}k_{1}^{2}-\frac{\mu_{23}}{\mu_{12}}\mathcal{E}}\right)\nonumber \\
 & +\frac{1}{\pi k_{1}}\int_{k_{1}-k_{F}}^{k_{1}+k_{F}}dp_{3}\,p_{3}\nonumber \\
 & \qquad\times\ln\left(\frac{p_{3}^{2}+k_{1}p_{3}+\frac{\mu_{23}}{\mu_{12}}k_{1}^{2}-\frac{\mu_{23}}{\mu_{12}}\mathcal{E}}{\frac{1}{2}p_{3}^{2}+(\frac{\mu_{23}}{\mu_{12}}-\frac{1}{2})k_{1}^{2}+\frac{1}{2}k_{F}^{2}-\frac{\mu_{23}}{\mu_{12}}\mathcal{E}}\right)\nonumber \\
 & +\frac{1}{\pi k_{1}}\lim_{\Lambda_{2}\rightarrow\infty}\int_{k_{1}+k_{F}}^{\Lambda_{2}}dp_{3}\,p_{3}\nonumber \\
 & \qquad\times\ln\left(\frac{p_{3}^{2}+k_{1}p_{3}+\frac{\mu_{23}}{\mu_{12}}k_{1}^{2}-\frac{\mu_{23}}{\mu_{12}}\mathcal{E}}{p_{3}^{2}-k_{1}p_{3}+\frac{\mu_{23}}{\mu_{12}}k_{1}^{2}-\frac{\mu_{23}}{\mu_{12}}\mathcal{E}}\right).\label{Omega_second_case_1}
\end{align}
We calculate each integral and use Eq. (\ref{s wave scattering length}),
which results in

\begin{align}
\Omega_{23}= & \frac{1}{a_{23}}-\frac{2k_{F}}{\pi}-\frac{2\sqrt{\kappa}}{\pi}\left[\arctan\left(\frac{\frac{1}{2}k_{1}-k_{F}}{\sqrt{\kappa}}\right)+\right.\nonumber \\
 & \left.-\arctan\left(\frac{\frac{1}{2}k_{1}+k_{F}}{\sqrt{\kappa}}\right)+\frac{\pi}{2}\right]+\nonumber \\
 & +\frac{1}{\pi k_{1}}\left((\frac{\mu_{23}}{\mu_{12}}-\frac{1}{2})k_{1}^{2}+k_{F}^{2}-\frac{\mu_{23}}{\mu_{12}}\mathcal{E}\right)\nonumber \\
 & \times\ln\left(\frac{\frac{\mu_{23}}{\mu_{12}}k_{1}^{2}+k_{F}k_{1}+k_{F}^{2}-\frac{\mu_{23}}{\mu_{12}}\mathcal{E}}{\frac{\mu_{23}}{\mu_{12}}k_{1}^{2}-k_{F}k_{1}+k_{F}^{2}-\frac{\mu_{23}}{\mu_{12}}\mathcal{E}}\right).\label{Omega_second_case_result}
\end{align}

\setcounter{equation}{0} \renewcommand{\theequation}{D\arabic{equation}}

\section*{appendix d. calculation of the function {\large{}$\Omega_{12}$} }

For a noninteracting mixture, $g_{23}=0$, the system of the integral
Eqs. (\ref{Main Eq 1}) and (\ref{Main Eq 2}) reduces to

\begin{multline}
\left[\frac{1}{g_{12}}+\int\frac{d^{3}\mathbf{p}_{3}}{(2\pi)^{3}}K_{1}(\mathbf{k}_{2},\mathbf{p}_{3};E)\right]F_{2}(\mathbf{k}_{2})\\
=-\int\frac{d^{3}\tilde{\mathbf{p}}_{3}}{(2\pi)^{3}}\tilde{K}_{1}(\mathbf{k}_{2},\tilde{\mathbf{p}}_{3};E)F_{2}(\tilde{\mathbf{p}}_{3}),\label{main integral Eq for g23 Zero}
\end{multline}
where the integral kernels $K_{1}$ and $\tilde{K}_{1}$ are given
by Eqs. (\ref{K1}) and (\ref{K1_tilde}), respectively. The cutoff
function $\theta_{\Lambda_{1}}(-\mathbf{k}_{2}-\mathbf{p}_{3})$,
which appears in $K_{1}$, imposes an upper bound, $u_{\mathrm{max}}$,
on the angle between the two momenta $\mathbf{k}_{2}$ and $\mathbf{p}_{3}$,
$u\equiv\cos\vartheta_{\mathbf{p}_{3},\mathbf{k}_{2}}$:

\begin{equation}
u_{\mathrm{max}}=\min_{p_{3}}\left(1,\frac{\Lambda_{1}^{2}-k_{2}^{2}-p_{3}^{2}}{2k_{2}p_{3}}\right)\rightarrow1\;\text{as }\Lambda_{1}\rightarrow\infty.\label{u_max}
\end{equation}
Next, without loss of generality we assume that $\mathbf{p}_{3}=p_{3}\mathbf{e}_{z}$,
where $\mathbf{e}_{z}$ is the unit vector in the direction of the
$z$-axis. For contact interactions and \emph{s}-wave symmetry of
the states we write Eq. (\ref{main integral Eq for g23 Zero}) as
Eq. (\ref{main Eq for g23 Zero}), where

\begin{align}
\Omega_{12}\equiv & \Omega_{12}(a_{12},k_{2};k_{F},\mathcal{E})\nonumber \\
\equiv & \frac{4\pi\hbar^{2}}{2\mu_{12}g_{12}}+\frac{1}{2\pi\frac{\mu_{12}}{m_{1}}k_{2}}\lim_{\Lambda_{2}\rightarrow\infty}\int_{k_{F}}^{\Lambda_{2}}dp_{3}\nonumber \\
 & \times p_{3}\ln\left(\frac{p_{3}^{2}+\frac{2\mu_{12}}{m_{1}}k_{2}p_{3}+k_{2}^{2}-\mathcal{E}}{p_{3}^{2}-\frac{2\mu_{12}}{m_{1}}k_{2}p_{3}+k_{2}^{2}-\mathcal{E}}\right).\label{Omega_12_V1}
\end{align}
Here, $\mathcal{E}=2\mu_{12}E/\hbar^{2}$ and $E$ is the energy of
the three-body system. We calculate the integral (\ref{Omega_12_V1}),
and use Eq. (\ref{s wave scattering length}) to obtain:

\begin{align}
\Omega_{12}= & \frac{1}{a_{12}}-\frac{k_{F}}{\pi}+\frac{\sqrt{\eta}}{\pi}\left[\arctan\left(\frac{\frac{\mu_{12}}{m_{1}}k_{2}+k_{F}}{\sqrt{\eta}}\right)\right.\nonumber \\
 & \left.-\arctan\left(\frac{\frac{\mu_{12}}{m_{1}}k_{2}-k_{F}}{\sqrt{\eta}}\right)-\pi\right]+\frac{1}{4\pi\frac{\mu_{12}}{m_{1}}k_{2}}\nonumber \\
 & \times\left[\left(2(\frac{\mu_{12}}{m_{1}})^{2}-1\right)k_{2}^{2}-k_{F}^{2}+\mathcal{E}\right]\nonumber \\
 & \times\ln\left(\frac{k_{2}^{2}+\frac{2\mu_{12}}{m_{1}}k_{F}k_{2}+k_{F}^{2}-\mathcal{E}}{k_{2}^{2}-\frac{2\mu_{12}}{m_{1}}k_{F}k_{2}+k_{F}^{2}-\mathcal{E}}\right),\label{Omega_12_result}
\end{align}
where $\eta\equiv[1-(\mu/m_{1})^{2}]k_{2}^{2}-\mathcal{E}$. 

\setcounter{equation}{0} \renewcommand{\theequation}{E\arabic{equation}}

\section*{appendix e. numerical solution of the system of integral eqs. (\ref{Main Eq 1})
and (\ref{Main Eq 2})}

Recall that we only consider the isotropic solutions of Eqs. (\ref{Main Eq 1})
and (\ref{Main Eq 2}), i.e., $F_{i}(\mathbf{k})=F_{i}(k)$. To solve
the system of the two coupled integral Eqs. (\ref{Main Eq 1}) and
(\ref{Main Eq 2}) we replace the three-dimensional integrals over
momentum by the absolute value of each momentum. Next, we calculate
the two functions $\Omega_{23}$ and $\Omega_{12}$ analytically;
see Appendices C and D. The analytical results reveal the lowest-energy
dimer state and the two-body bound-state continuum. We solve the coupled
Eqs. (\ref{Main Eq 1}) and (\ref{Main Eq 2}) for a given three-body
parameter $\Lambda\gg k_{F}$. For that, we discretize the integral
ranges on the grid points $\{x_{j}^{(N)}\}$, $j=1,2,\ldots,N$, that
are the sets of zeros of the Legendre polynomials $P_{N}(x)$. We
approximate each integral by a truncated sum that is weighted by $w_{j}^{(N)}$:

\begin{equation}
w_{j}^{(N)}=\frac{2}{1-[x_{j}^{(N)}]^{2}}\frac{1}{[P'_{N}(x_{j}^{(N)})]^{2}},\label{Gauss-Legendre weights}
\end{equation}
where $P'_{N}(x)=dP_{N}(x)/dx$ \cite{Gaussian_quadrature_1,Gaussian_quadrature_2}.
This choice is the so-called Gauss-Legendre quadrature rule, supporting
the highest order of accuracy among the other quadrature rules \cite{Gaussian_quadrature_1}.

We apply the Gauss-Legendre quadrature rule on each integral and construct
a matrix equation analog to an integral equation. For given values
of $E$ below the threshold energy (\ref{threshold energy}), we calculate
the eigenvalues resulting in the corresponding values of the $s$-wave
scattering lengths. The unknown functions $F_{1}$ and $F_{2}$ will
be obtained as the eigenvectors of the matrix equations.

\setcounter{equation}{0} \renewcommand{\theequation}{F\arabic{equation}}

\section*{appendix f. derivation of eq. (\ref{a_12 critical trimer})}

The atoms ``1'' and ``2'' interact attractively via contact interactions
according to Eq. (\ref{Cooper interaction operator}). We follow Appendix
B and rewrite the Schr{\"o}dinger equation describing the pair-12
in terms of the relative momentum, $\mathbf{p}_{12}\equiv(\mu_{12}/m_{1})\mathbf{k}_{1}-(1-\mu_{12}/m_{1})\mathbf{k}_{2}$,
and the total momentum, $\mathbf{P}_{12}\equiv\mathbf{k}_{1}+\mathbf{k}_{2}$,
as 

\begin{equation}
\frac{4\pi\hbar^{2}}{2\mu_{12}g_{12}}=-4\pi\int\frac{d^{3}\mathbf{p}_{12}}{(2\pi)^{3}}\,\frac{1}{p_{12}^{2}+\frac{\mu_{12}}{m_{1}}(1-\frac{\mu_{12}}{m_{1}})P_{12}^{2}-\mathcal{E}_{12}},\label{integral Eq for two-body 12}
\end{equation}
where $\mathcal{E}_{12}=2\mu_{12}E_{12}/\hbar^{2}$, $E_{12}$ is
the energy of the pair-12, and $\mu_{12}$ is a reduced mass, $1/\mu_{12}=1/m_{1}+1/m_{2}$.
The Fermi sea demands a constraint on the momentum of the atom ``2'',
$k_{2}>k_{F}$, which in terms of the relative and total momenta reads
$|\frac{\mu_{12}}{m_{1}}\mathbf{P}_{12}-\mathbf{p}_{12}|>k_{F}$.
This constraint imposes an upper bound on $\cos\vartheta_{\mathbf{p}_{12},\mathbf{P}_{12}}$.
Without loss of generality we assume that $\mathbf{P}_{12}=P_{12}\mathbf{e}_{z}$,
where $\mathbf{e}_{z}$ is the unit vector in the direction of the
$z$-axis. 

\begin{figure}[b]
\includegraphics[width=1\columnwidth]{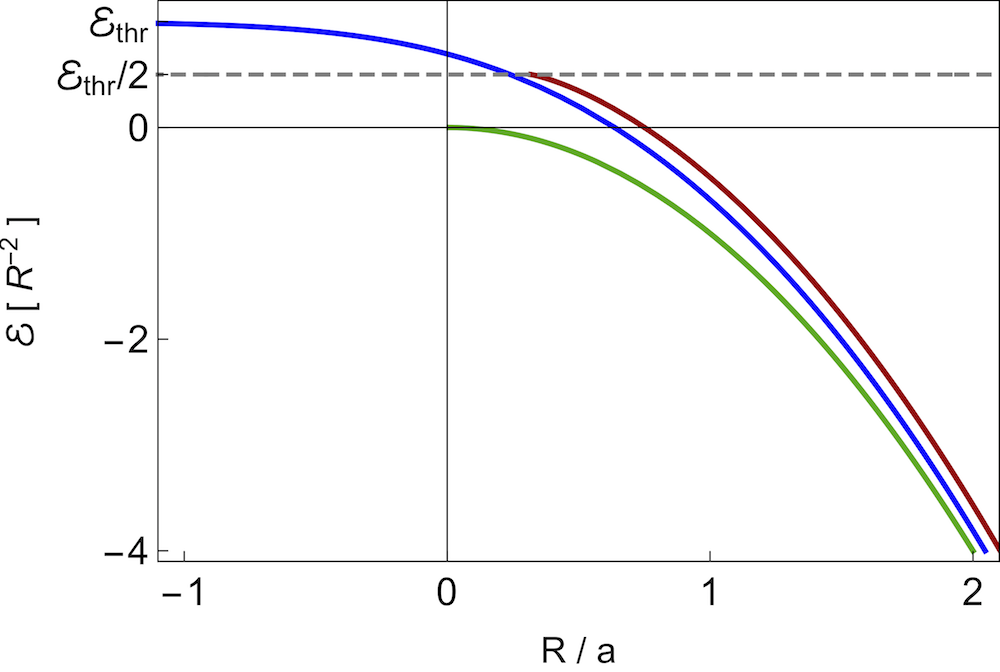}

\caption{Energy $\mathcal{E}=2\mu E/\hbar^{2}$ in units of $R^{-2}$ vs $R/a$
for two equal-mass atoms with a reduced mass $\mu$ and the \emph{s}-wave
scattering length $a$, where $R$ denotes an arbitrary length scale.
The green curve is the result in vacuum, $k_{F}=0$, given by Eq.
(\ref{first dimer in vacuum}). The blue curve shows the result of
a Cooper pair with vanishing total momentum described by Eq. (\ref{Cooper pair solution}),
where both atoms are immersed in an inert Fermi sea with the Fermi
momentum $k_{F}R=1$. The red curve is the result for a pair with
the total momentum $k_{F}$, where one atom is in vacuum and the other
is subject to an inert Fermi sea with the Fermi momentum $k_{F}R=1$;
cf. Eqs. (\ref{a12 for pair-12 with P _1 result}) and (\ref{a12 for pair-12 with P -2 result}).
The gray dashed lines show $\mathcal{E}_{\mathrm{thr}}$ and $\mathcal{E}_{\mathrm{thr}}/2$,
where $\mathcal{E}_{\mathrm{thr}}=2\mu E_{\mathrm{thr}}/\hbar^{2}=k_{F}^{2}$.}

\label{Fig10}
\end{figure}

To solve Eq. (\ref{integral Eq for two-body 12}) analytically, we
assume \emph{s}-wave symmetry of the states and consider two cases.
For $P_{12}\leqslant(\mu_{12}/m_{1})^{-1}k_{F}$ we have:

\begin{align}
\frac{4\pi\hbar^{2}}{2\mu_{12}g_{12}}= & \frac{-1}{\frac{2\mu_{12}}{m_{1}}\pi P_{12}}\int_{k_{F}-\frac{\mu_{12}}{m_{1}}P_{12}}^{k_{F}+\frac{\mu_{12}}{m_{1}}P_{12}}dp_{12}\,p_{12}\nonumber \\
 & \times\frac{p_{12}^{2}+(\frac{\mu_{12}}{m_{1}})^{2}P_{12}^{2}-k_{F}^{2}}{p_{12}^{2}+\frac{\mu_{12}}{m_{1}}(1-\frac{\mu_{12}}{m_{1}})P_{12}^{2}-\mathcal{E}_{12}}\nonumber \\
 & -\frac{1}{\pi}\int_{k_{F}-\frac{\mu_{12}}{m_{1}}P_{12}}^{k_{F}+\frac{\mu_{12}}{m_{1}}P_{12}}dp_{12}\,p_{12}^{2}\nonumber \\
 & \times\frac{1}{p_{12}^{2}+\frac{\mu_{12}}{m_{1}}(1-\frac{\mu_{12}}{m_{1}})P_{12}^{2}-\mathcal{E}_{12}}\nonumber \\
 & -\frac{2}{\pi}\int_{k_{F}+\frac{\mu_{12}}{m_{1}}P_{12}}^{\Lambda_{2}}dp_{12}\,p_{12}^{2}\nonumber \\
 & \times\frac{1}{p_{12}^{2}+\frac{\mu_{12}}{m_{1}}(1-\frac{\mu_{12}}{m_{1}})P_{12}^{2}-\mathcal{E}_{12}}.\label{a12 for pair-12 with P _1 v1}
\end{align}
We calculate each integral, take the limit $\Lambda_{2}\rightarrow\infty$,
and use Eq. (\ref{s wave scattering length}). The result is

\begin{align}
\frac{1}{a_{12}}= & \frac{k_{F}}{\pi}-\frac{1}{\pi}\sqrt{\varrho}\left[\arctan\left(\frac{k_{F}-\frac{\mu_{12}}{m_{1}}P_{12}}{\sqrt{\varrho}}\right)\right.\nonumber \\
 & \left.+\arctan\left(\frac{k_{F}+\frac{\mu_{12}}{m_{1}}P_{12}}{\sqrt{\varrho}}\right)-\pi\right]+\frac{1}{4\pi\frac{\mu_{12}}{m_{1}}P_{12}}\nonumber \\
 & \times\left(\frac{\mu_{12}}{m_{1}}(\frac{2\mu_{12}}{m_{1}}-1)P_{12}^{2}-k_{F}^{2}+\mathcal{E}_{12}\right)\nonumber \\
 & \times\ln\left(\frac{\frac{\mu_{12}}{m_{1}}P_{12}^{2}-\frac{2\mu_{12}}{m_{1}}k_{F}P_{12}+k_{F}^{2}-\mathcal{E}_{12}}{\frac{\mu_{12}}{m_{1}}P_{12}^{2}+\frac{2\mu_{12}}{m_{1}}k_{F}P_{12}+k_{F}^{2}-\mathcal{E}_{12}}\right),\label{a12 for pair-12 with P _1 result}
\end{align}
where $\varrho\equiv\frac{\mu_{12}}{m_{1}}(1-\frac{\mu_{12}}{m_{1}})P_{12}^{2}-\mathcal{E}_{12}$
. 

For $P_{12}\geqslant(\mu_{12}/m_{1})^{-1}k_{F}$ we have:

\begin{align}
\frac{4\pi\hbar^{2}}{2\mu_{12}g_{12}}= & -\frac{2}{\pi}\int_{0}^{\frac{\mu_{12}}{m_{1}}P_{12}-k_{F}}dp_{12}\,p_{12}^{2}\nonumber \\
 & \times\frac{1}{p_{12}^{2}+\frac{\mu_{12}}{m_{1}}(1-\frac{\mu_{12}}{m_{1}})P_{12}^{2}-\mathcal{E}_{12}}\nonumber \\
 & -\frac{1}{\frac{2\mu_{12}}{m_{1}}\pi P_{12}}\int_{\frac{\mu_{12}}{m_{1}}P_{12}-k_{F}}^{\frac{\mu_{12}}{m_{1}}P_{12}+k_{F}}dp_{12}\,p_{12}\nonumber \\
 & \times\frac{p_{12}^{2}+(\frac{\mu_{12}}{m_{1}})^{2}P_{12}^{2}-k_{F}^{2}}{p_{12}^{2}+\frac{\mu}{m_{1}}(1-\frac{\mu}{m_{1}})P_{12}^{2}-\mathcal{E}_{12}}\nonumber \\
 & -\frac{1}{\pi}\int_{\frac{\mu_{12}}{m_{1}}P_{12}-k_{F}}^{\frac{\mu_{12}}{m_{1}}P_{12}+k_{F}}dp_{12}\,p_{12}^{2}\nonumber \\
 & \times\frac{1}{p_{12}^{2}+\frac{\mu_{12}}{m_{1}}(1-\frac{\mu_{12}}{m_{1}})P_{12}^{2}-\mathcal{E}_{12}}\nonumber \\
 & -\frac{2}{\pi}\int_{\frac{\mu_{12}}{m_{1}}P_{12}+k_{F}}^{\Lambda_{2}}dp_{12}\,p_{12}^{2}\nonumber \\
 & \times\frac{1}{p_{12}^{2}+\frac{\mu_{12}}{m_{1}}(1-\frac{\mu_{12}}{m_{1}})P_{12}^{2}-\mathcal{E}_{12}}.\label{a12 for pair-12 with P _2 v1}
\end{align}
We calculate each integral, take the limit $\Lambda_{2}\rightarrow\infty$,
use Eq. (\ref{s wave scattering length}), and arrive at:

\begin{align}
\frac{1}{a_{12}}= & \frac{k_{F}}{\pi}+\frac{1}{\pi}\sqrt{\varrho}\left[\arctan\left(\frac{\frac{\mu_{12}}{m_{1}}P_{12}-k_{F}}{\sqrt{\varrho}}\right)\right.\nonumber \\
 & \left.-\arctan\left(\frac{\frac{\mu_{12}}{m_{1}}P_{12}+k_{F}}{\sqrt{\varrho}}\right)+\pi\right]+\frac{1}{4\pi\frac{\mu_{12}}{m_{1}}P_{12}}\nonumber \\
 & \times\left(\frac{\mu_{12}}{m_{1}}(\frac{2\mu_{12}}{m_{1}}-1)P_{12}^{2}-k_{F}^{2}+\mathcal{E}_{12}\right)\nonumber \\
 & \times\ln\left(\frac{\frac{\mu_{12}}{m_{1}}P_{12}^{2}-\frac{2\mu_{12}}{m_{1}}k_{F}P_{12}+k_{F}^{2}-\mathcal{E}_{12}}{\frac{\mu_{12}}{m_{1}}P_{12}^{2}+\frac{2\mu_{12}}{m_{1}}k_{F}P_{12}+k_{F}^{2}-\mathcal{E}_{12}}\right);\label{a12 for pair-12 with P -2 result}
\end{align}
see Fig. \ref{Fig10}.

As discussed in the text, for $m_{2}/m_{1}\gg1$ we estimate the onset
of a highest-energy excited three-body bound state at zero energy
by calculating the onset of the lowest-energy pair-12. To do that,
we expand Eq. (\ref{a12 for pair-12 with P _1 result}) or Eq. (\ref{a12 for pair-12 with P -2 result})
for $m_{2}/m_{1}\gg1$, as $\mathcal{E}_{12}\rightarrow0$ and $P_{12}\rightarrow(\frac{\mu_{12}}{m_{1}})^{-1}k_{F}$,
which results in Eq. (\ref{a_12 critical trimer}).

\setcounter{equation}{0} \renewcommand{\theequation}{G\arabic{equation}}

\section*{appendix g. calculation of the parameter {\large{}{\lowercase{$s_{0}$}}}}

The Efimov scaling factor is $\lambda=\exp(\pi/|s_{0}|)$, where the
effect of the mass ratio $m_{2}/m_{1}$ is described by the parameter
$s_{0}$. For \emph{s}-wave symmetry of the states, if we have a system
of three species only with two-resonantly interacting pairs, then
$s_{0}$ is the purely imaginary root of the transcendental equation

\begin{equation}
\cos\left(\frac{\pi}{2}s_{0}\right)=\frac{2}{\sin2\vartheta}\frac{\sin(\vartheta s_{0})}{s_{0}},\label{Eq for s0 for two interacting pairs}
\end{equation}
where $\vartheta=\arcsin[(m_{2}/m_{1})/(1+m_{2}/m_{1})]$, $\vartheta\in[0,\pi/2]$.
If all three species are resonantly interacting, we obtain $s_{0}$
as the purely imaginary root of the equation

\begin{multline}
\left[\cos\left(\frac{\pi}{2}s_{0}\right)-\frac{2}{\sin2\vartheta}\frac{\sin(\vartheta s_{0})}{s_{0}}\right]\cos\left(\frac{\pi}{2}s_{0}\right)\\
=\frac{8}{\sin^{2}2\gamma}\frac{\sin^{2}(\gamma s_{0})}{s_{0}^{2}},\label{Eq for s0 for three interacting pairs}
\end{multline}
where $\gamma=\arcsin\{\sqrt{(m_{1}/m_{2})/[2(1+m_{2}/m_{1})]}\}$,
$\gamma\in[0,\pi/4]$. For a proof, see Ref. \cite{Pascal_review_paper}.


\begin{thebibliography}{10}
\bibitem{momentum_transfer}By ``momentum transfer'' we mean the
difference of the in-state and out-state momenta of a particle; see,
e.g., Ref. \cite{Scattering_Book}.
\bibitem{excited_trimers}For Efimov states in vacuum, $k_{F}=0$,
by increasing the mass ratio $m_{2}/m_{1}$, the Efimov scaling factor
$\exp(\pi/|s_{0}|)$ decreases; cf. Appendix G. This implies that
for a given value of $\Lambda$, increasing $m_{2}/m_{1}$ leads to
more excited Efimov states. For $k_{F}\neq0$ the Efimov spectrum
is deformed near unitarity and the Efimov scaling factor does not
hold anymore; nevertheless, as we increase $m_{2}/m_{1}$, excited
three-body bound states appear.
\bibitem{Error}An estimation of the numerical error occurred in our
calculations is discussed in Appendix E.
\end{thebibliography}
\end{document}